\documentclass{aa}  

\usepackage{graphicx}
\usepackage{txfonts}
\usepackage[colorlinks=true,allcolors=blue]{hyperref}
\newcommand{\crat}[1]{$^{12}\text{C}/^{13}\text{C}#1$}
\newcommand{\feh}[1]{$\text{[Fe/H]}#1$}
\usepackage{color}
\definecolor{darkred}{rgb}{0.75,0.,0.5}

\definecolor{darkgreen}{rgb}{0.0,0.205,0.}

\begin{document}

   \title{Explaining the $^{12}\text{C}/^{13}\text{C}$ ratio in the Galactic halo: the contribution from shell mergers in primordial massive stars}   \titlerunning{The $^{12}\text{C}/^{13}\text{C}$ ratio in the Galactic halo}


   \author{
          F. Rizzuti\inst{1,2,3}
          \and
          G. Cescutti\inst{1,2,3}
          \and
          P. Molaro\inst{2,4}
          \and
          L. Roberti\inst{5,6,7}
          \and
          A. Chieffi\inst{8,9,10}
          \and
          M. Limongi\inst{7,9,11}
          \and
          L. Magrini\inst{12}
          \and
          F. Matteucci\inst{1,2,3}
          }

\institute{Dipartimento di Fisica, Università degli Studi di Trieste, via Tiepolo 11, I-34143 Trieste, Italy\\
              \email{federico.rizzuti@inaf.it}
         \and
             INAF, Osservatorio Astronomico di Trieste, via Tiepolo 11, I-34143 Trieste, Italy
        \and
        INFN, Sezione di Trieste, via Valerio 2, I-34134 Trieste, Italy
        \and
        Institute for Fundamental Physics of the Universe, via Beirut, 2, I-34151 Trieste, Italy
        \and
        Konkoly Observatory, Research Centre for Astronomy and Earth Sciences, HUN-REN, Konkoly Thege Miklós út 15-17, 1121 Budapest, Hungary
        \and
        CSFK, MTA Centre of Excellence, Konkoly Thege Miklós út 15-17, 1121 Budapest, Hungary
        \and
        INAF, Osservatorio Astronomico di Roma, Via Frascati 33, I-00040 Monteporzio Catone, Italy
        \and
        INAF/IAPS, Via Fosso del Cavaliere 100, I-00133 Roma, Italy
        \and
        INFN, Sezione di Perugia, via A. Pascoli s/n, I-06125 Perugia, Italy
        \and
        Monash Centre for Astrophysics (MoCA), School of Mathematical Sciences, Monash University, Victoria 3800, Australia
        \and
        Kavli IPMU (WPI), The University of Tokyo, Kashiwa, 277-8583 Chiba, Japan
        \and
        INAF, Osservatorio Astrofisico di Arcetri, Largo E. Fermi 5, I-50125 Firenze, Italy
}

   \date{Received ; accepted }

 
  \abstract
   {Recent campaigns of observations have provided new measurements of the carbon isotopes in the most metal poor stars of the Galaxy. These stars are so metal-poor that they could only have been enriched by one or few generations of massive progenitors. However, explaining the primary production of $^{13}$C and the low \crat{} ratio measured in these stars is challenging.}
   {Making use of the most up-to-date models for zero-metal and low metallicity stars, we investigate the possible sources of $^{13}$C at low metallicity and verify whether massive stars could be the sole responsible for the \crat{} ratio observed in halo stars.}
   {We employ the stochastic model for Galactic chemical evolution \texttt{GEMS} to reproduce the evolution of CNO elements and \crat{} ratio, including the enrichment from rotating massive stars, some of which show the occurrence of H-He shell mergers.}
   {We find that stars without H-He shell mergers do not produce enough $^{13}$C to be compatible with the observations. Instead, the primary production by shell mergers and later ejection during the supernova explosion can explain 30 < \crat{<100}. The observations are best reproduced assuming a large frequency of shell mergers. The \crat{<30} can only be reproduced assuming an outer layer ejection and no explosion, but requiring a larger production of $^{12}$C and $^{13}$C. }
   {Zero-metal and low-metallicity spinstars with H-He shell mergers appear as the most plausible scenario to explain the low \crat{} ratio in CEMP-no stars. 
   The entire range of \crat{} values can be explained by assuming that some stars fully explode while others only eject their outer layers. Shell mergers should be also more frequent and productive, which is allowed by the current uncertainties in the treatment of convection in stellar modelling.}


   \maketitle
%
\section{Introduction}
The old and metal poor stars that compose the Galactic halo possess critical information about the initial phases of galaxy evolution \citep{1951ApJ...114...52C, 1984ApJ...285..622B, 1985AJ.....90.2089B}. Indeed, while zero-metal stars (also known as Population III stars) have never been directly observed, presumably because stellar formation is biased towards short-lived massive stars at lower metallicity, many of the oldest stars in the Galactic halo have so low metal content that they are believed to have formed from the explosion of Pop III stars as type-II supernovae, therefore locking in the new stellar atmosphere the chemical composition processed by only one generation of primordial stars. These second-generation stars with subsolar masses survived until today, representing some of the oldest and most metal-poor stars that can be observed in the Milky Way \citep[see][]{2023A&A...669L...4A}.\\
Recent spectroscopic surveys are providing thousands of high-resolution spectra for metal-poor halo stars \citep{2000AJ....120.1579Y, 2007PASA...24....1K, 2009AJ....137.4377Y, 2012RAA....12..735D, 2013A&A...560A..71C, 2017MNRAS.471.2587S, 2022ApJS..259...60R}. Considering that direct detection of Pop III stars from primordial galaxies at high redshift is still unviable, the abundance measurements of metal-poor halo stars represent the only way of constraining the properties and nucleosynthesis of the first generation of stars.\\
Ever since the first surveys of metal-poor halo stars \citep{1985AJ.....90.2089B} and the first spectroscopic follow-ups  \citep{1990A&A...228..426M, 1990A&A...236L...5M}, a large fraction of stars was found to be highly enriched in carbon \citep{1992AJ....103.1987B} compared to the solar value ([C/Fe] > 0.7). These carbon-enhanced metal-poor (CEMP) stars represent the totality of stars at metallicity \feh{<-5}, and only two stars out of 14 known at \feh{<-4.5} have normal solar abundance. Some CEMP stars also show the presence of neutron-capture elements in their atmospheres, while others do not: when [Ba/Fe] > 1 these are called CEMP-s stars, while if [Ba/Fe] < 0 they are CEMP-no stars. CEMP-no stars outnumber CEMP-s below \feh{<-2.3} \citep{2018ApJ...861..146Y}. For CEMP-no stars, it is believed that the large carbon abundance comes from the gas in which the star formed, previously enriched by Pop III stars, but in order to explain CEMP-s stars it is normally invoked the presence of an asymptotic giant branch (AGB) companion that accretes both carbon and s-process elements onto the CEMP star. The two classes have also different levels of carbon \citep{2013A&A...552A.107S, 2015A&A...579A..28B}, and almost all CEMP-s stars are found in binary systems \citep{2005ApJ...625..825L, 2014MNRAS.441.1217S}.\\
Explaining the chemical composition and large carbon enrichment of CEMP-no stars has always been challenging. Carbon is a primary element that is produced in both massive and low-intermediate mass stars (LIMS). Considering that CEMP-no stars are the result of a few or even just one progenitor, their composition must be explained either by a single source abundant in carbon but scarce in iron \citep{2003Natur.422..871U}, or multiple sources whose ejecta are enriched in light elements and pollute only a small region \citep{2003Natur.422..834B, 2003PASA...20..324C}. Overall, the mechanisms that determine the explosion of massive stars as type-II supernovae (SNe II) are still not completely understood, given the interplay of several complex phenomena 
\citep{2015MNRAS.448.2141M, 2015ApJ...806..275P, 2019ApJ...870....2C, 2021Natur.589...29B, 2024Univ...10..148B}. It has been constrained from observations \citep[since][]{Sollerman_1998,Turatto_1998} that SNe II must possess a distribution of explosion energies, in some cases ejecting very low masses of iron. These so-called `faint SNe' represent an excellent candidate to explain the nature of CEMP-no stars, since they eject little Fe but large amounts of CNO elements \citep{2003Natur.422..834B, 2003PASA...20..324C, 2003Natur.422..871U,2005ApJ...619..427U}.  \\
CEMP-no stars are also interesting sites to explore the behaviour of the isotopic ratio \crat. Normally, $^{13}$C is produced in H-burning regions by the CNO cycle, but it is later destroyed during the evolution of the star. A possible way of producing and preserving superficial $^{13}$C is in AGB stars due to hot bottom burning or hot-third dredge up that leads to H-burning at the bottom of the convective envelope  \citep{1981A&A....94..175R, 1999ARA&A..37..239B, 2009ApJ...696..797C, 2011ApJS..197...17C, 2010MNRAS.403.1413K}. However, a low \crat{} ratio, therefore large amounts of $^{13}$C, has been often observed in CEMP-no stars \citep{Spite, 2021A&A...652A..97S}, which are so metal-poor that only massive stars could have contributed to their enrichment. This requires that $^{13}$C is produced as a primary element at low metallicity, as it is the case for $^{14}$N: in fact, \cite{1986MNRAS.221..911M} first showed that nitrogen requires also a primary production by massive stars, so the same is expected to be valid for $^{13}$C \citep[see also][]{2003MNRAS.342..185R}.\\
A possible path for $^{13}$C production is when H- and He-burning regions get in contact at late times in the stellar evolution. Mixing episodes between H- and He-burning regions have been reported in stellar modelling for a long time, especially for massive zero-metal stars \citep{1982ASIC...90...79W, 2010ApJ...724..341H, 2012ApJS..199...38L}. The subsequent introduction of rotation in stellar modelling \citep{2002A&A...390..561M, 2008A&A...489..685E} has even enhanced these effects, due to the rotation-induced mixing. Motivated by observations of low \crat{} in unmixed stars \citep{Spite}, \cite{2007A&A...461..571H} has shown that a low \crat{} ratio can be obtained in massive fast rotators ($v_\text{ini} = 600$ - 800 km s$^{-1}$) down to $Z=10^{-8}$, thanks to the rotational mixing that transports carbon from the He-burning core up to the tail of the H-burning shell, therefore producing primary $^{13}$C and $^{14}$N. This has been confirmed and further explored by \cite{2017A&A...605A..63C}. A similar production of $^{13}$C also in absence of rotation has been shown to be possible in massive zero-metal stars \citep{2012ApJS..199...38L, 2021MNRAS.500.2685C}, due to convection-induced mixing and proton ingestion into He-burning regions \citep[see also][]{2011ApJ...727...89H}. The two mechanisms, i.e.\ rotation- and convection-induced mixing, can also coexist \citep{Roberti_2024}. \\
These occurrences are sometimes called `shell mergers', in cases when the H- and He-burning shells are sufficiently close to each other that convection overcomes the inter-shell barrier and material can mix between the two shells, sometimes merging into a new single convective region. The H-rich material is entrained and burns with the C of the He-burning shell, producing $^{13}$C and $^{14}$N, and then it is transported close to the surface by the strong convection of the shell merger, later enriching the ISM. The occurrence of shell mergers in massive stars and their frequency are still largely debated, but recent indications from both 1D and multi-D stellar models show that they may occur quite often \citep{2014ApJ...783...10S, 2018MNRAS.473.1695C, Roberti_2024, 2024MNRAS.533..687R}. H-He shell mergers are more common at low metallicity due to the smaller entropy jump between the shells and partial activation of the 3$\alpha$ process, making the shells closer both in mass and in radius \citep[see e.g.][]{Roberti_2024}.\\
Overall, rotation is believed to play a key role in Pop III stars. Stellar modelling has been showing \citep{2002A&A...390..561M, 2007A&A...461..571H, 2016MNRAS.456.1803F, Limongi_2018} that massive stars at low metallicity are expected to have higher rotation velocity, due to their lower opacity, larger compactness, and lower mass loss. Fast metal-poor rotators have been used in studies of Galactic archaeology to explain both light and heavy element production at low metallicity \cite[since][]{2006A&A...449L..27C, 2008A&A...479L...9C, 2010A&A...515A.102C, 2013A&A...553A..51C}, thanks to the rotation-induced mixing that affects the core size and transports the products closer to the surface. These same mechanisms may also have a key role for explaining the carbon enhancement of CEMP-no stars.\\
Indeed, Galactic archaeology represents a precious method to validate stellar modelling results, by relating the yields from stellar models to the actual observations. \cite{2008A&A...479L...9C} have been able to show first, using the yields of \cite{2007A&A...461..571H}, that a very low \crat{\sim30} can be reached at low metallicity thanks to fast rotators employed in a chemical evolution model. More recently, \cite{2017MNRAS.470..401R, 2019MNRAS.490.2838R} and \cite{2018MNRAS.476.3432P} have confirmed and further investigated these results employing more recent grids of rotating massive stars \citep[see also][]{2022A&ARv..30....7R}. The recent measurements of \crat{} in increasingly lower-metallicity stars \citep{2022A&A...668A..86A,Molaro} further motivate the research of their production sites.\\
In this paper, we present the results from a stochastic chemical evolution model of the Galactic halo including the nucleosynthesis from zero-metal and metal-poor stars, in order to explain the production of CNO elements and the \crat\ ratio recently measured in halo stars. \\
The paper is organised as follows: in Section 2, we list the adopted observations; in Section 3, we present the chemical evolution model; in Section 4, the nucleosynthesis prescriptions are discussed; in Section 5, the results are presented; and in Section 6, the conclusions are drawn.

\section{Observations}\label{sec:obs}
\begin{table*}
\centering
\footnotesize
\caption{Observational data used for this work: source, number of giants (defined as $\log g<3.2$), number of dwarfs ($\log g\geq3.2$), method of analysis, spectral resolution, and abundances used.}\label{tab:obs}
\begin{tabular}{cccccc}
\hline\\[-0.7em]
Source&n\_giants&n\_dwarfs&method&spectral resolution&abundances\\
\\[-0.7em]\hline\\[-0.7em]
\cite{2014ApJ...797...21P} (compilation) &335$^\text{a}$&158&1D LTE&35 000 - 60 000&C, N, Sr, Ba\\[0.2em]
{\cite{Molaro}} (compilation)&27&24&1D LTE&32 000 - 140 000&$^{12}$C, $^{13}$C, Sr, Ba\\[0.2em]
{\cite{2004A&A...421..649I}}&0&31&1D LTE&35 000 - 50 000&N, O\\[0.2em]
{\cite{2005A&A...430..655S}}&16$^\text{b}$&0&1D LTE$^\text{c}$&47 000&N, O\\[0.2em]
{\cite{2008ApJ...681.1524L}}&16&12&1D LTE&60 000&C, N, O\\[0.2em]
{\cite{2009A&A...500.1143F}}&0&43&1D NLTE&60 000&C, O\\[0.2em]
{\cite{2013ApJ...762...25N}}&15&6&1D LTE&40 000&O\\[0.2em]
{\cite{2016ApJ...833..225Z}}&0&11&1D NLTE&60 000&C, O\\[0.2em]
{\cite{CescuttiMince, 2024A&A...686A.295F}}&41&0&1D LTE&65 000 - 115 000&Sr, Ba\\
\\[-0.7em]\hline\\[-0.7em]
\multicolumn{6}{l}{\begin{minipage}{0.85\textwidth}$^\text{a}$ An evolutionary correction has been applied to the C abundances \citep{2014ApJ...797...21P}.\\
$^\text{b}$ Only unmixed giants have been selected, using the surface abundance of Li as a diagnostic of the mixing \citep{2005A&A...430..655S}.\\
$^\text{c}$ Nitrogen and oxygen have been corrected for 3D effects \citep{2005A&A...430..655S}.\end{minipage}}
\end{tabular}
\end{table*}

In this paper, we reproduce the chemical evolution of the isotopes of carbon and the elements nitrogen, oxygen, barium, and strontium for stars in the Galactic halo. Given that these abundances are measured in stellar atmospheres with different methods, we rely here on multiple works that include measurements of the most metal poor stars in the literature. \\
We present in Table \ref{tab:obs} the observations we employ in this work, listing the number of giants and dwarfs, with details of their analysis. We are aware that the original surface abundances of a star can be modified at the end of its lifetime, due to internal mixing processes during the red giant phase such as first dredge-up and thermohaline mixing \citep{2023A&A...676A..19M, 2024MNRAS.530..761M, 2024arXiv240805039N}. This affects mainly the light elements, while the heavy elements are mostly left untouched. In order to avoid this source of uncertainty, we only include dwarf stars in our CNO observation sample, selected as having $\log g\geq3.2$. Exceptions are studies that provide evolutionary corrections for C in giants \citep{2014ApJ...797...21P}, or that use Li to detect unmixed giants \citep{2005A&A...430..655S}. For Sr and Ba instead we include giants. We also present in the appendix the Kiel diagram of the observations (Fig.~\ref{fig:iso}). \\ 
In particular, \cite{2014ApJ...797...21P} provide a compilation of observations for C, N, Sr, and Ba (see references within), including an evolutionary correction for C to recover its original surface abundance in red giant branch (RGB) stars. \\
Oxygen is particularly challenging to measure in stars at low metallicity. We include measurements for C, N, and O from the works of \cite{2004A&A...421..649I, 2005A&A...430..655S, 2008ApJ...681.1524L, 2009A&A...500.1143F, 2013ApJ...762...25N, 2016ApJ...833..225Z}. The observations of \cite{2009A&A...500.1143F} and \cite{2016ApJ...833..225Z} have been obtained with non-LTE analysis. \cite{2005A&A...430..655S} have corrected their observations for 3D effects by $-0.40$ for nitrogen (NH band) and by $-0.23$ for oxygen ([O I] line).\\
Sr and Ba at intermediate metallicity come from the MINCE project by \cite{CescuttiMince,2024A&A...686A.295F}. \\
For $^{12}$C and $^{13}$C, we employ the compilation of stars presented in the recent work of \cite{Molaro}, which also includes new measurements for six stars among the most metal-poor known in the halo, in addition to a list of measurements from previous works\footnote{\cite{1997ApJ...489L.169N, 1998A&A...332..672B, 2015A&A...579A..28B, 2018A&A...612A..65B, 2002ApJ...567.1166A, 2006A&A...459..125S, 2010A&A...509A..93M, 2012A&A...548A..34A, 2013ApJ...762...25N, 2014Natur.506..463K, 2015ApJ...810L..27F, 2019ApJ...871..146F, 2015ApJ...807..173H, 2016A&A...595L...6C, 2017PASJ...69...24M, 2018ApJ...852L..20A, 2019ApJ...874L..21A, 2019MNRAS.488L.109N, 2020ApJ...889L..13G, 2021A&A...652A..97S, 2022A&A...668A..86A, 2023A&A...669L...4A}.}. We also include the dwarf stars HD 140283 from \cite{Spite,2021A&A...652A..97S} and HE 0007-1832 from \cite{2004ApJ...612.1107C}. It is important to underline that 6 stars in our sample with \crat{<30} are dwarfs, therefore their low \crat{} is not an effect of internal mixing: CS 22887-048, CS 22945-017, CS 22956-028, CS 22958-042, G77-61, and HE 0007-1832.\\
We adopted in our models the solar abundances of \cite{2009ARA&A..47..481A}. This is not always the case for the observations listed above, especially the works before 2009, which used various prescriptions (see each work). Overall, these very small differences do not invalidate the comparison of the models to the observations.

\section{Chemical evolution model}
For this work, we have developed a new version of the stochastic model first presented in \cite{2010A&A...515A.102C}, which was based on the inhomogeneous model of \cite{2008A&A...481..691C} and on the homogeneous model of \cite{2008A&A...479L...9C}. The stochastic model has been introduced to reproduce the scatter visible in the observations of neutron-capture elements at low metallicity, which can be explained by the randomly distributed physical properties of the stars \citep{2013A&A...553A..51C, 2014A&A...565A..51C, 2016A&A...595A..91C, Rizzuti2021}. In this sense, chemical evolution models that take this dispersion into account have additional degrees of freedom that can be used to further constrain the stellar properties \citep[see also][]{1999ApJ...519L..63T, 1999ApJ...511L..33I, 2001ApJ...547..217T, 2002A&A...388..842A, 2004A&A...416..997A, 2005A&A...436..879K, 2015MNRAS.452.1970W}. \\
We have now further developed the stochastic model by adapting the code to parallel computing, using domain decomposition through the MPI (Message-Passing-Interface) library for interprocess communication. This allowed us to dramatically increase the number of volumes used to simulate the Galactic halo. We also expanded the list of isotopes and added to the code new sources of nucleosynthesis for zero-metal and low metallicity stars (see Section \ref{sec:nuc}). We named this new release of the stochastic model \texttt{GEMS} (Galactic Evolution via Montecarlo Sampling). 
\\
We recall here that the model is constructed to reproduce the chemical history of the Galactic halo, therefore it runs for 1 Gyr after the formation of the Galaxy. To reproduce the inhomogeneities in the chemical composition, the simulation domain is divided into subvolumes, which are considered independent. Each cell is assumed to be a cube of volume $8\times10^6$ pc$^3$ as in \cite{2015A&A...577A.139C}, in order to take into account the distance covered by the ejecta of supernovae \citep[$\sim50$ pc,][]{1998ApJ...500...95T}, so to avoid exchange of material between cells. Taking a volume too large would reduce the stochasticity and produce more homogeneous results. On the other hand, the metallicity that we are trying to reproduce is so low, that there are very few events that are responsible for the enrichment before reaching higher metallicity. For this reason, we had to drastically increase the number of volumes in the simulation, going from 100 of \cite{2008A&A...481..691C} and 1000 of \cite{2016A&A...595A..91C}, to 10 000 volumes in this work, thanks to the parallel computing of the \texttt{GEMS} code. Despite the large number of cells, the total volume of our simulations still represents only less than 0.1 per cent of the total volume of the Galactic halo. \\
The model we employ here has the same specifics as described in \cite{Rizzuti2021}. In particular, we recall that only one infall episode is assumed for the halo, accreting gas of primordial composition according to a Gaussian distribution \citep{2008A&A...479L...9C}:
\begin{equation}
\dot{G}(t)_\text{inf} = \dfrac{\Sigma_\text{h}\ A}{\tau \sqrt{2\pi}}\ \text{e}^{-(t-t_0)^2/2\tau^2}
\end{equation}
where $t_0$ is 100 Myr, $\tau$ is 50 Myr, and the normalization is given by $\Sigma_\text{h}=80\ \text{M}_{\odot}\ \text{pc}^{-2}$ and $A$ the surface of each cell, so that $\Sigma_\text{h}\ A=3.2\times10^6\ \text{M}_{\odot}$. \\
The initial mass function (IMF) $\phi(m)$ adopted here is the one from \cite{1986FCPh...11....1S}. \\
The star formation rate (SFR), in units of $\text{M}_{\odot}\ \text{Gyr}^{-1}$, is defined as
\begin{equation}
\psi(t) =\nu\ \Sigma_\text{h}\ A\ \left(\dfrac{G_\text{gas}(t)}{\Sigma_\text{h}\ A}\right)^k
\end{equation}
where $\nu$ is 1.4 Gyr$^{-1}$, $k$ equals 1.5, and $G_\text{gas}(t)$ is the amount of gas inside the volume in M$_{\odot}$.
\\Finally, we take into account the Galactic wind that takes gas out of each cell, with a rate that is proportional to the star formation rate:
\begin{equation}
\dot{G}(t)_\text{out} =\omega\ \psi(t)
\end{equation}
according to a constant $\omega$ set equal to 8 for all chemical species \citep{2008A&A...479L...9C}.\\
The evolution proceeds in the following way. In order to ensure stochasticity, in each cell at each time-step a certain mass of gas, according to the SFR, is converted into stars, which are randomly extracted with mass between 0.1 and 100 M$_\odot$. In this way, each cell has the same amount of stars in mass but with a different distribution. After the stars are born, their evolution is followed until they die, assuming the stellar lifetimes of \cite{1989A&A...210..155M}.

\section{Nucleosynthesis prescriptions}\label{sec:nuc}
As we mentioned in the introduction, in order to explain the CNO abundances and \crat\ ratio observed in halo stars, we need to assume specific nucleosynthesis sources. In the \texttt{GEMS} code, the productive sources considered for the chemical enrichment are LIMS and rotating massive stars, and their endpoints as AGB stars, type-Ia SNe, core-collapse SNe, and neutron star mergers. Although all these sources are present in the model, in order to explain the low-metallicity observations at \feh{<-2}, only sources with short time-scales are predominant, in particular massive stars and their supernova explosions. We list below the nucleosynthesis assumptions we include in the model.

\subsection{Faint supernovae and iron}
\begin{figure}
   \centering
   \includegraphics[width=\columnwidth]{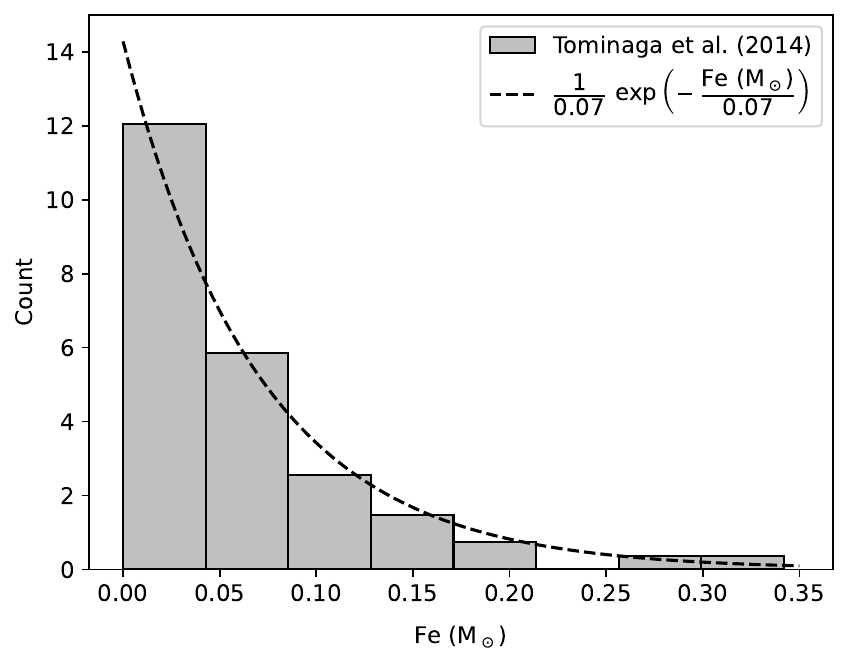}
      \caption{Number distribution of ejected Fe mass from low-metallicity SNe, estimations from \cite{Tominaga_2014} (histogram) and exponential fitting (dashed line) used to reproduce faint SNe.}\label{fig:tomin}
   \end{figure}

Over the years, the observations of CEMP stars with very low iron content but high carbon suggested the existence of stars that explode as so-called `faint SNe' \citep{2003Natur.422..834B, 2003PASA...20..324C, 2003Natur.422..871U, 2005ApJ...619..427U, 2007ApJ...660..516T, 2014ApJ...785...98T, 2013ARA&A..51..457N}. These events have been introduced to reproduce the abundance patterns and other observational data from peculiar stars and SN remnants \citep{Sollerman_1998,Turatto_1998}. It is assumed that a large amount of mixing and fallback in the inner layers prevents the star from ejecting most of its Fe, but still allows it to eject the outer layers rich in CNO elements. Compared to standard core-collapse SN that have an explosion energy of $\sim10^{51}$ erg, 3D SN simulations are also able to reproduce low-luminosity explosions, reaching energies $<10^{50}$ erg \citep{10.1093/mnras/staa1691, 10.1093/mnras/stac1518}. \\
Since we know very little of the physics and distribution of these faint SNe, we formulate here some prescriptions based on observational constraints. \cite{2016A&A...595A..91C}, when implementing faint SNe in their chemical evolution model, assumed a linear distribution of Fe yields from zero to 0.2 M$_\odot$, therefore with an expected value of 0.1 M$_\odot$. We use here a different approach: \cite{2014ApJ...785...98T} constrained the nucleosynthesis of faint SNe by producing SN models that reproduce the abundance patterns of 48 stars with \feh{< -3.5}. It can be noticed (see Fig.~\ref{fig:tomin}) that the distribution of ejected Fe mass roughly follows an exponential probability distribution:
\begin{equation}\label{eq:exp}
f(x;\beta)=
\begin{cases}
\dfrac{1}{\beta}\ \text{e}^{-x/\beta} & \quad x \ge 0 \\
0 & \quad x < 0
\end{cases}
\end{equation} 
that can be fitted so that the expected value E$[X]=\beta$ is around 0.07, as expected from core-collapse SN. \\
In this study, we assume that core-collapse SNe eject an amount of Fe mass following distribution (\ref{eq:exp}), with $\beta$ the Fe yield given by the stellar evolution model. The yields assumed for Fe come from the same models of massive stars that produce the CNO elements (see next section). Note that with this approach, faint SNe always eject some iron, even if only a few percent of a regular supernova. This means that most of the other elements are also ejected, in particular the CNO elements produced close to the surface; this is necessary in order to reproduce CEMP-no stars with abundant C and low but non-zero Fe. The scenario of a faint supernova is different from the one of a `failed' supernova, in which despite the unsuccessful explosion it is still possible that only the outer layers are ejected, enriched in light elements, but no iron.

\subsection{Stellar yields for C, N, O}
Carbon, nitrogen and oxygen can be produced by stars of different masses; therefore, their time-scale of production is dependent on the lifetime and mass of the star. For the metallicity range we are most interested in (\feh{< -2}), the main producers are the massive stars ($M>8$ M$_\odot$). Non-rotating stellar yields traditionally used in chemical evolution studies \citep[e.g.][]{1995ApJS..101..181W} have been shown to underestimate the production of both light and heavy elements \citep[e.g.][]{2016A&A...595A..91C, 2022A&ARv..30....7R, 2024A&A...691A.284R}. Instead, rotation has been shown to have a critical impact on the stellar nucleosynthesis \citep{2002A&A...390..561M, 2007A&A...461..571H}, especially at low metallicity, where stars are more compact due to their lower opacity and tend to rotate faster. Additionally, rotating massive stars are the only way to explain the production of primary $^{13}$C and $^{14}$N observed in low metallicity stars \citep{2006A&A...449L..27C,2008A&A...479L...9C}. \\
Recently, several works have produced grids of models for rotating massive stars at low metallicity, providing stellar yields to be used for Galactic archaeology \citep{2007A&A...461..571H, 2012A&A...538L...2F, 2016MNRAS.456.1803F, Limongi_2018, Roberti_2024}. In particular, the stellar models produced with the \texttt{FRANEC} code \citep{2013ApJ...764...21C} represent one of the most complete grids of models in the literature. Among their strengths are their treatment of rotation, evolution until the pre-SN stage, and explosive nucleosynthesis. \cite{Limongi_2018} provide a grid of nine masses between 13 - 120 M$_\odot$, four metallicities \feh{= 0, -1, -2,} and --3, and three rotations with an initial equatorial velocity of 0 km s$^{-1}$ (non rotating), 150 km s$^{-1}$, and 300 km s$^{-1}$. This grid of models has proved to be extremely interesting for the recent developments in chemical evolution studies \citep{2018MNRAS.476.3432P, 2023MNRAS.523.2126P, 2019MNRAS.490.2838R, Rizzuti2019, 2020MNRAS.498.1252G, 2022IAUS..366...63K, 2023MNRAS.523.2974M, 2024A&A...691A.284R}. \\
The yields of \cite{Limongi_2018} have been also implemented in the stochastic model of \cite{Rizzuti2021}, which is at the basis of the \texttt{GEMS} code presented here. Although the basic implementation remains the same, there are important differences that it is worth underlining here. \cite{Limongi_2018} provide different sets of yields, depending on the physical assumptions. Differently from \cite{Rizzuti2021}, we used here their Set R (recommended), calculated using the mixing and fallback technique \citep{2003Natur.422..871U, 2005ApJ...619..427U} assuming that each star ejects 0.07 M$_\odot$ of $^{56}$Ni, but stars > 25 M$_\odot$ fully collapse to a black hole, therefore enriching the ISM only of wind during their lifetime. This assumption is consistent with observations of red supergiants in the Local Group with masses up to 25 M$_\odot$ \citep{Smartt}, and theoretical studies failing to explode stars above 25 - 30 M$_\odot$ \citep{2016ApJ...821...38S, 2016MNRAS.460..742M, 2020ApJ...890...43C, 2023ApJ...949...17B}, however we recognise that this threshold of explodability is still an open problem and subject to uncertainty. Having only wind contribution from stars > 25 M$_\odot$ and not the explosion decreases the enrichment in almost all elements. Considering that fast rotation enhances the nucleosynthesis instead, the net effect is that low-metallicity stars are required to rotate faster to produce the same enrichment. The exact rotation velocity function we assume is described below.\\
To reproduce the observations below \feh{=-3}, which is the lowest metallicity in the \cite{Limongi_2018} grid of stellar models, we extended the grid with the recent work of \cite{Roberti_2024}, who also employed the \texttt{FRANEC} code, providing 15 and 25 M$_\odot$ models at [Fe/H] =  $-4, -5,$ and zero metal ($Z=0$) for a wide range of rotation velocities. Although partial, this grid proves to be crucial for our investigation. We complement the grid in the following way: stars between 8 - 15 M$_\odot$ have the same nucleosynthesis as 15 M$_\odot$, between 15 - 25 M$_\odot$ are linearly interpolated, and above 25 M$_\odot$ we assume no production, considering that we expect only a wind contribution, which is always orders of magnitude lower than the explosive one. It should be noticed that most 15 M$_\odot$ models have C-O shell mergers, which affect the abundances in the inner layers \citep[see][]{Roberti_2024}.\\
Finally, the \texttt{GEMS} code includes also the contribution from LIMS, whose impact is visible in the model results only from around \feh{> -2}. The yields have been assumed from stellar models obtained with the \texttt{FRUITY} code \citep{2009ApJ...696..797C, 2011ApJS..197...17C, 2015ApJS..219...40C} for non-rotating stars between 1.3 - 6 M$_\odot$.

\subsection{Stellar yields for neutron capture elements}
Massive stars are important producers of both light and heavy elements; in particular, they can also produce trans-iron nuclei when rotation is taken into account, thanks to the s-process enabled by rotational mixing between the H-shell and He-core \citep{2012A&A...538L...2F, 2016MNRAS.456.1803F, 2013ApJ...764...21C, Limongi_2018}. The present work is not focused on studying the evolution of neutron-capture elements, since this has already been done by \cite{Rizzuti2021} with very similar assumptions. However, in changing the implementation of massive stars to reproduce the CNO elements, we are also changing the production of s-elements, since they come from the same sites. A holistic approach to Galactic archaeology would require the same chemical evolution model to reproduce the evolution of multiple elements under the same assumptions, rather than fine-tuning every time the individual elements. For these reasons, we show in this paper also the predicted evolution of heavy elements strontium and barium, after the model has been fine-tuned to explain the CNO isotopes. In this way, we intend to show that the model does not have to be calibrated again to explain the evolution of different elements, but a convergence of results over the same assumptions is achievable.\\
A detailed description of the neutron-capture nucleosynthesis and its implementation in the stochastic model has already been presented in \cite{Rizzuti2021}. We recall here that the production of heavy elements requires multiple sources, namely rotating massive stars (s-process) and neutron star mergers or magneto-rotationally driven SNe (r-process) at lower metallicity, and AGB stars (s-process) at higher metallicity. The stellar yields we employ come from the same sources described in the previous sections; however, for LIMS the yields from non-rotating models tend to overproduce the neutron capture elements at solar metallicity, while rotating models have too little production instead. Therefore, as already done in \cite{Rizzuti2019, Rizzuti2021}, we divide the non-rotating yields by a factor of 2, so that the abundance of neutron capture elements observed at solar metallicity can be reproduced. \\
Neutron star mergers are implemented in the \texttt{GEMS} code to reproduce the r-process production of heavy elements. Their implementation is described in \cite{2015A&A...577A.139C}, in particular, the productive mass range is 9 - 50 M$_\odot$, the fraction of binary systems is 0.018, and the coalescence time-scale is fixed to 1 Myr, consistently with what found by \cite{2014MNRAS.438.2177M}. The assumption of a constant time-delay is simplistic, considering that more complex prescriptions have been developed in the recent years \citep{2019MNRAS.486.2896S, 2021MNRAS.500.1071M, 2021MNRAS.503....1C}. However, it can still reproduce the observations, and the question of the productive sites for the r-process is still largely debated today. In particular, our models show that below \feh{<-3} the s-process from massive stars dominates the production of Sr and Ba, while from \feh{>-3} a mixture of s- and r-process is responsible \citep[see also][]{2014A&A...565A..51C}. 

\subsection{Rotation velocity in massive stars}
To reproduce neutron-capture elements measured in halo stars, \cite{Rizzuti2021} have calibrated the rotational velocity of low-metallicity massive stars through observations assuming a distribution in rotational velocities. Given that only three rotations are provided in the \cite{Limongi_2018} grid, \cite{Rizzuti2021} used interpolation between different velocities. Here, we improve the study by extending the grid of stellar models below \feh{=-3}. A new calibration would be required at this point, similar to what was done in \cite{Rizzuti2021}, for the new yields and assumptions. However, the stellar grid that we use below \feh{=-3} \citep[i.e.][]{Roberti_2024} is still incomplete in mass. For this reason, we make here the simplest assumptions to avoid additional sources of uncertainty. \\
We assume the same rotation velocity for stars with the same mass and metallicity so that we do not interpolate between stellar models with different rotations, given the possible non-linearity of stellar physics. Therefore, we give all massive stars the same rotation velocity depending on their metallicity:
\begin{equation}\label{eq:vel}
\text{velocity([Fe/H])}=
\begin{cases}
300 \text{ km s$^{-1}$} & \quad \text{[Fe/H]} \le -3 \\
150 \text{ km s$^{-1}$} & \quad \text{[Fe/H]} > -3
\end{cases}
\end{equation} 
This distribution is inspired by the assumptions and results of \cite{2018MNRAS.476.3432P} and \cite{Rizzuti2021}, but considering a faster rotation to compensate for the enrichment only by winds above 25 M$_\odot$. We are interested here only in metal-poor stars below \feh{<-2}, therefore we do not decrease the rotation even further towards solar metallicity, as discussed in \cite{Rizzuti2019}. 

\section{Results}

\subsection{The \crat{} ratio in stellar models}
\begin{figure*}
   \centering
   \includegraphics[width=\hsize]{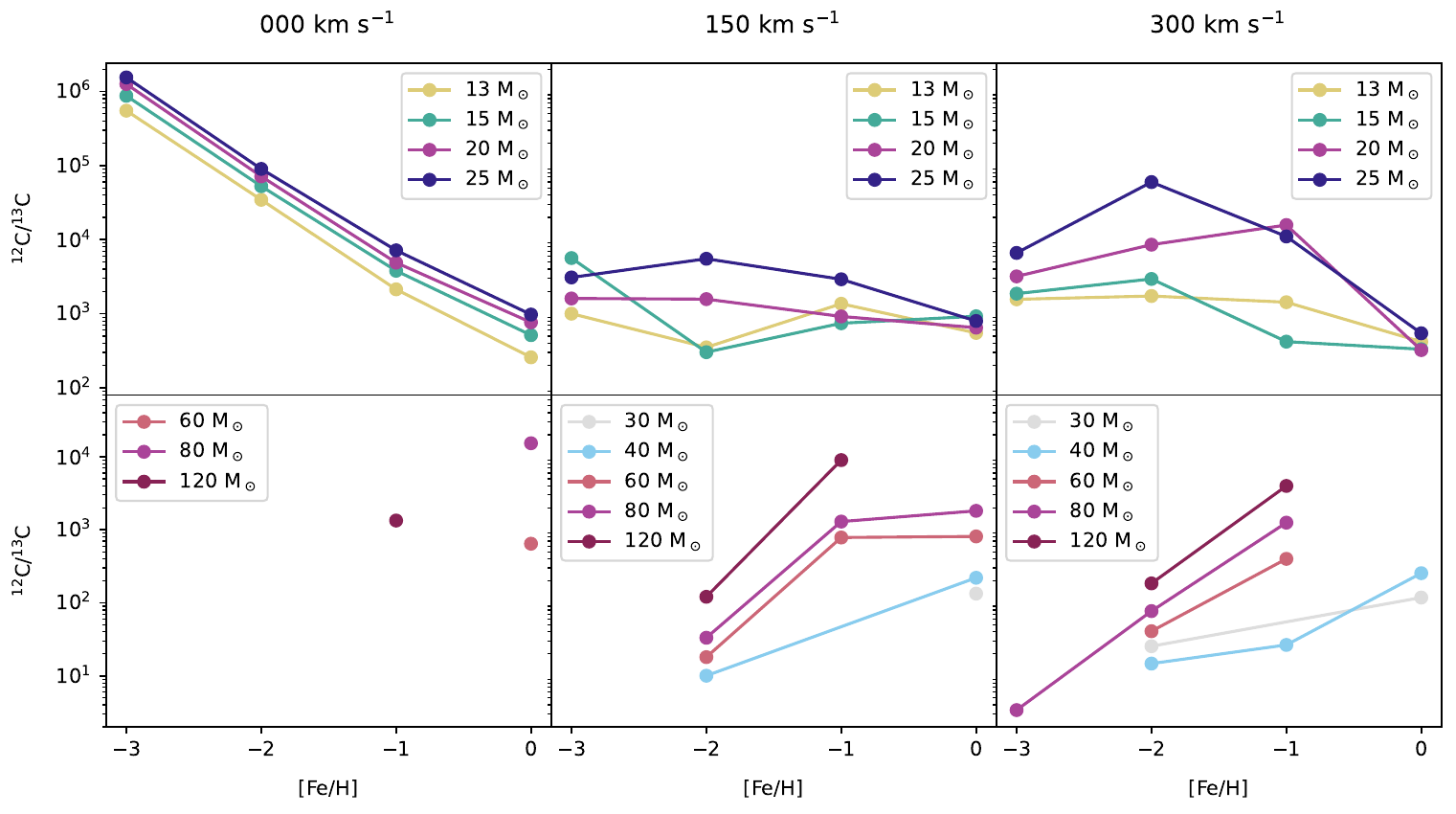}
      \caption{\crat{} ratio in the stellar models of \cite{Limongi_2018}, for stars that contribute via both wind and explosion (first row) or only wind (second row), for the three initial rotation velocities 0 (left), 150 (centre) and 300 (right) km s$^{-1}$. Where dots are missing the models do not predict enrichment of $^{12}$C or $^{13}$C, or both.}
         \label{fig:limyields}
   \end{figure*}

\begin{figure*}
   \centering
   \includegraphics[width=\hsize]{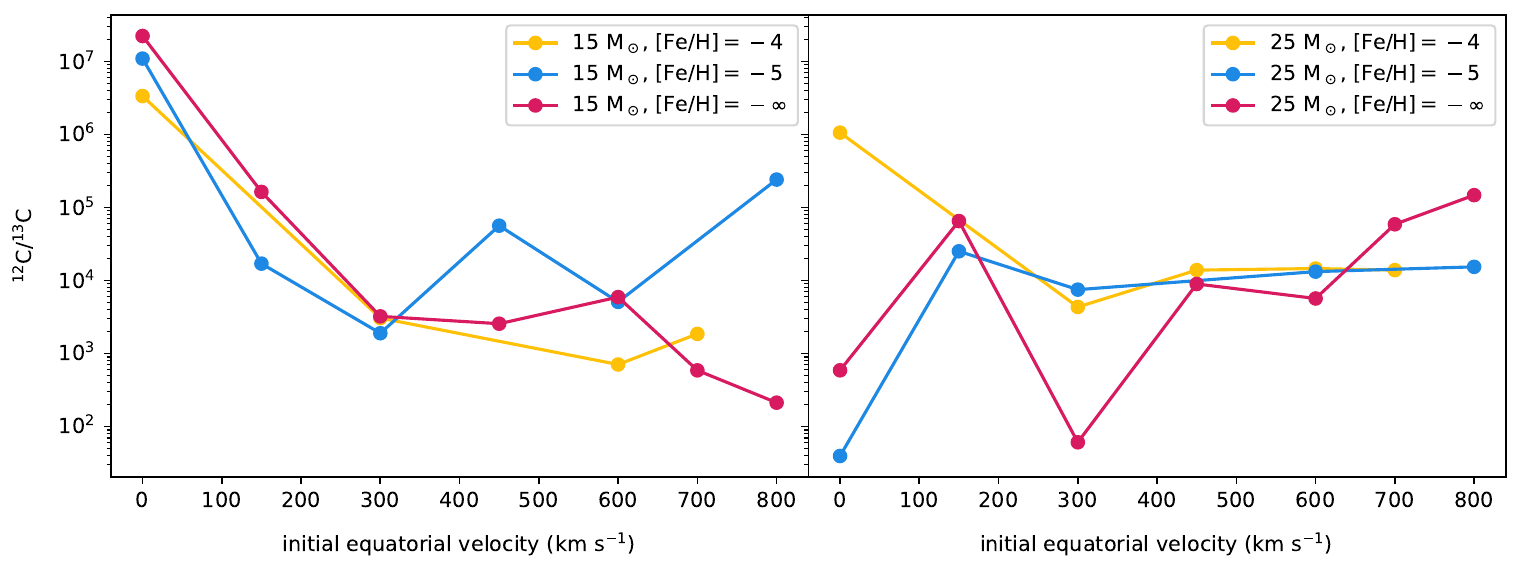}
      \caption{\crat{} ratio in the stellar models of \cite{Roberti_2024}, for 15 (left) and $25\text{ M}_\odot$ (right) stars, as a function of the star initial equatorial velocity in km s$^{-1}$. Different colours correspond to different metallicities: [Fe/H] $= -4, -5$ and $-\infty$ (zero metals).}
         \label{fig:robyields}
   \end{figure*}

\begin{figure}
   \centering
   \includegraphics[width=\hsize]{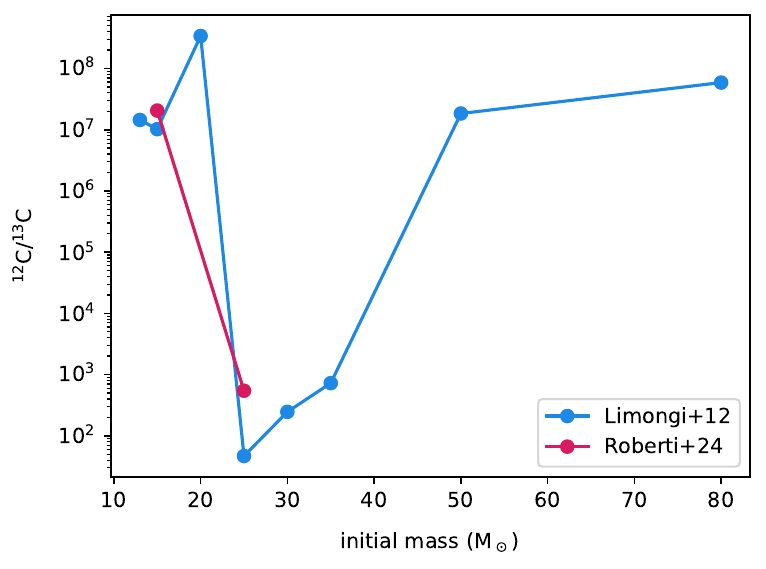}
      \caption{\crat{} ratio in the stellar models of \cite{2012ApJS..199...38L} compared to \cite{Roberti_2024}, as a function of the stellar mass, for non-rotating zero-metal stars.}
         \label{fig:lim12}
   \end{figure}

\begin{figure}
   \centering
   \includegraphics[width=\hsize]{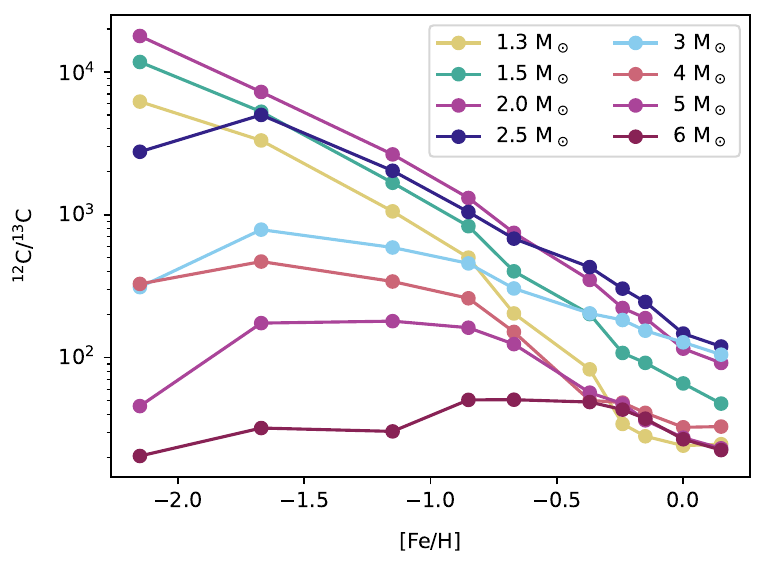}
      \caption{\crat{} ratio in the LIMS models of \cite{2009ApJ...696..797C, 2011ApJS..197...17C, 2015ApJS..219...40C} as a function of metallicity, for different non-rotating stellar masses.}
         \label{fig:criyields}
   \end{figure}

As a first step towards explaining the \crat{} ratio measured in halo stars, we present here a detailed analysis of the stellar yields produced by some of the most recent studies on stellar modelling. The abundances observed in extremely metal-poor halo stars (\feh{\leq -3}) suggest that the gas that formed them had been enriched only by a few generations of progenitors, if not directly from zero-metal stars. This means that the contribution to the chemistry of those stars can only come from massive stars, which are the first to die and enrich the ISM. Therefore, we need to identify the possible stellar models of massive stars that produce the correct amount of $^{12}$C and $^{13}$C actually observed in metal-poor stars. We remind that all the massive star models we analyse in this work provide explosive yields, assuming that both inner and outer layers of the star are ejected; we treat the case of partial ejection in Section \ref{sec:dep}. All \crat{} ratios we compute and show in this paper are abundances by number.\\
\cite{Limongi_2018} have developed a set of models with different mass, metallicity, and rotation velocity (see Section \ref{sec:nuc}). We compute and show in Fig.~\ref{fig:limyields} the \crat{} ratio for all these models, grouped by mass and rotation velocity. In particular, the first row of Fig.~\ref{fig:limyields} shows the masses $13\text{ - 25 M}_\odot$, that in \cite{Limongi_2018} contribute via both wind and explosion, while the $30\text{ - 120 M}_\odot$ stars in the second row only contribute via stellar wind.
The observations of halo stars presented in Section \ref{sec:obs} all measure a \crat{} ratio below 100. We can clearly see from Fig.~\ref{fig:limyields} that stellar models between $13\text{ - 25 M}_\odot$ (first row) all predict a \crat{} ratio well above 100, due to their very small production of $^{13}$C. On the other hand, a few more massive stars that only enrich through the wind (Fig.~\ref{fig:limyields}, second row) can reach \crat{} < 100. However, the amount of $^{13}$C ejected is very small ($<10^{-3}$ M$_\odot$ for \feh{=-3}), and the birth of stars $>25\text{ M}_\odot$ is disfavoured compared to less massive stars, so their enrichment would soon be mixed with $^{12}$C-rich material from SN ejecta, and the \crat{} ratio would greatly increase.\\
Therefore, the \cite{Limongi_2018} stellar models down to \feh{=-3} cannot reproduce the \crat{} ratio measured from observations, which are mostly below \feh{<-3} and would require extrapolation of the yields at lower metallicity. This implies that a different mechanism of production for $^{13}$C is necessary at extremely low metallicity. \\
As we introduced in Section \ref{sec:nuc}, a new set of \texttt{FRANEC} models has been published by \cite{Roberti_2024}, exploring 15 and $25\text{ M}_\odot$ stars at metallicity [Fe/H] $= -4, -5$, and zero metals ($Z=0$), for many different rotation velocities. Despite its much smaller mass range, its extremely low metallicity and large rotation range make this grid ideal for investigating the nucleosynthesis of the first stars. Similarly to what we did before, we show in Fig.~\ref{fig:robyields} the \crat{} ratio as predicted by the models of \cite{Roberti_2024}, plotted as a function of the stellar initial equatorial velocity. 
It can be noticed that, despite some exceptions, generally a faster rotation corresponds to lower \crat{}, therefore a larger production of $^{13}$C, up to around $10^{-3}$ M$_\odot$. This is due to the `entanglement' effect described in \cite{Roberti_2024}: the rotation-induced instabilities encourage the exchange of matter between the He-core and the H-shell, resulting in a larger concentration of CNO products in the core and stable intershell, including $^{13}$C. Later on, He-core and He-shell burning completely destroy the $^{13}$C through ($\alpha,$ n). Indeed, we see from Fig.~\ref{fig:robyields} that most models lie above \crat{>10^3}, indicating that this mechanism does not leave much $^{13}$C. However, there are a few models that lie below \crat{<10^3}, with some of them even below \crat{<10^2}. This larger production of $^{13}$C is the result of another type of occurrence.\\
In \cite{Roberti_2024}, the non-rotating $25\text{ M}_\odot$ models at \feh{=-5,-\infty}, and the 300 km s$^{-1}$ rotating one at \feh{=-\infty} (the ones with lowest \crat{} in Fig.~\ref{fig:robyields}, left) present a merging of the outer H-burning shell with the He-burning shell immediately below. In particular, once the He-shell penetrates upwards into the H-shell, the rapid engulfment of protons heating at He-burning temperature produces large amounts of $^{13}$C and $^{14}$N. Part of these products, instead of being immediately burnt, are convectively transported outside the He-burning region to the outer layers of the shell merger, where they can survive. This mechanism explains the large $^{13}$C production (up to $10^{-2}$ M$_\odot$) and low \crat{} ratio. These are the yields that we employ in the chemical evolution model in order to explain the observed \crat{} ratio in halo stars.\\
To shed more light on the effects of H-He shell mergers on the $^{13}$C production, we analyze also the results of \cite{2012ApJS..199...38L}, who presented a set of non-rotating zero-metal models in the range 13 - 80 M$_\odot$. The aim is to draw a comparison with the grid of models by \cite{Roberti_2024}, although only for non-rotating massive stars. We show in Fig.~\ref{fig:lim12} the predicted \crat{} in \cite{2012ApJS..199...38L} versus \cite{Roberti_2024} as a function of the initial stellar mass.
Although \cite{Roberti_2024} has modelled only two masses, results seem compatible between the two studies. Only stars with mass between 25 - 35 M$_{\odot}$ exhibit the H-He shell merger described above, resulting in large production and expulsion of $^{13}$C. These results support our choice of including shell merging events in our chemical evolution model.
\\Finally, to interpret the behaviour of the model above \feh{=-2}, we also briefly present the nucleosynthesis sources that contribute at higher metallicity, i.e.\ LIMS. We plot in Fig.~\ref{fig:criyields} the \crat{} ratio from the non-rotating 1.3 - 6 M$_\odot$ stellar models of \cite{2009ApJ...696..797C, 2011ApJS..197...17C, 2015ApJS..219...40C} that we use in the \texttt{GEMS} code. We can immediately notice that the ratio spans a wide range across $\sim20\text{ - }20\,000$ at low metallicity, which gradually reduces going towards solar metallicity. Their impact on the chemical evolution is shown in the next sections. 

\subsection{Chemical evolution of the \crat{} ratio}\label{sec:13c}
\begin{figure}
   \centering
   \includegraphics[width=\hsize]{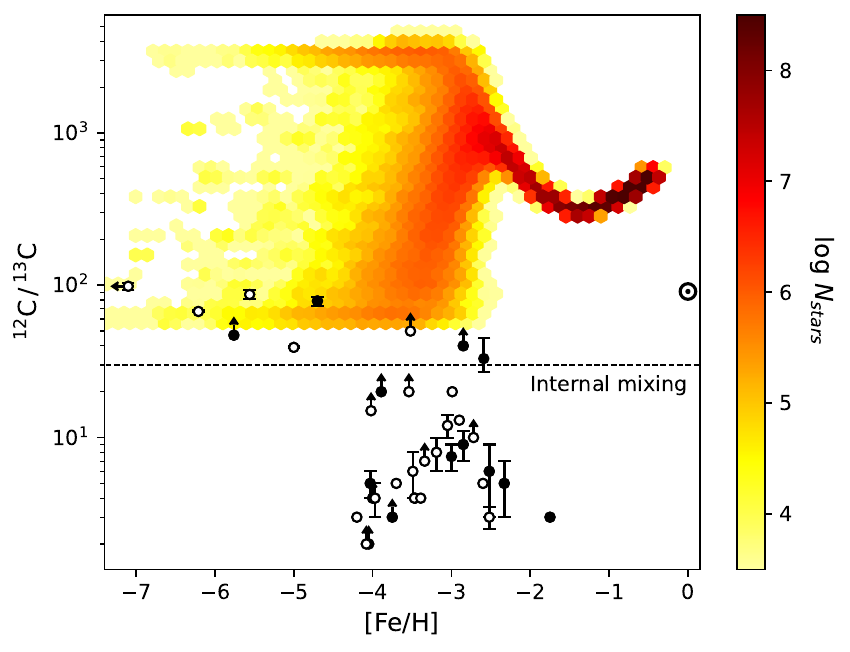}
      \caption{Isotopic ratio \crat{} versus [Fe/H]. The stochastic model is the colour map, with the number of simulated stars on a logarithmic scale. The dots are the observed halo stars (dwarfs in black, giants in white) as presented in \cite{Molaro}, complete of their possible error. The horizontal line \crat{ = 30} marks the evidence for internal mixing in giant stars \citep{Spite}. 
      }
         \label{fig:SFR}
   \end{figure}

We present here the results of the chemical evolution model for the \crat{} ratio, using the yields and assumptions described in Section \ref{sec:nuc}. In particular, we recall that massive stars rotate at 150 km s$^{-1}$ above [Fe/H] $>-3$ and 300 km s$^{-1}$ below [Fe/H] $\leq-3$, with the yields from \cite{Limongi_2018} for [Fe/H] $\geq-3$ and from \cite{Roberti_2024} for [Fe/H] $<-3$.\\
The model outcome is presented in Fig.~\ref{fig:SFR}. 
The behaviour of the model can be easily interpreted in light of the stellar yields we discussed above. Below \feh{<-4}, the model is dominated by the zero-metal massive stars: two long tracks extend towards the lowest metallicity, corresponding to the two zero-metal rotating models of 15 and $25\text{ M}_\odot$, plus their interpolation (no extrapolation is used, see Section \ref{sec:nuc}). Although the grid of yields is incomplete, we can see that the model is able to reproduce the observations at \feh{<-4} thanks to the presence of $25\text{ M}_\odot$ stars with predicted \crat{ = 56} due to shell merging. The model also extends towards much higher \crat{}, because the other stars produce small $^{13}$C and large \crat{> 10^3} due to their lack of shell merging. \\
Going towards \feh{>-4} the model rises sharply above \crat{> 10^3}, because the yields for massive stars by \cite{Limongi_2018} predict large \crat{} especially in the range $-3<$ \feh{<-2}. From \feh{>-2.5} the model starts decreasing again due to the contribution from AGB stars. As we showed in Fig.~\ref{fig:criyields}, this is due to the first 3 - 6 M$_\odot$ stars that die from around 60 Myr and \feh{=-3}, enriching the ISM of a \crat{} ratio below 300 due to hot bottom burning, strongly bringing down the ratio that was around $\sim 10^3$ due to massive stars. We confirm the contribution from massive AGB stars from the age-metallicity relation that we include in the appendix (Fig.~\ref{fig:age}). \\
From around \feh{>-1.5} also the lower-mass stars begin to die, and their \crat{} is high, so the model starts rising again. Nevertheless, we recognise that the scenario above \feh{>-2} is really complex, due to the different contributions from both low-mass and massive stars, while observations in this range are scarce and uncertain. Since we cannot use observations to constrain the producers of $^{13}$C, we prefer not to draw conclusions for this metallicity range. The solar value is \crat{=91\pm1.3} \citep{2003ApJ...598.1038G, 2013ApJ...765...46A}, however we recall that the version of the \texttt{GEMS} code we use here is for simulating the halo, therefore we do not expect the model to reproduce the solar value.\\
Finally, we notice that the model cannot reproduce the observations below \crat{<30} (indicated by a dashed line in Fig.~\ref{fig:SFR}). No stellar yield adopted in this study can produce enough $^{13}$C to reach such a low ratio. Although explaining the composition of these stars is difficult and still a matter of debate, there are possible explanations that involve the physics of the producers or of the host stars. We comment on this point later in Sec.~\ref{sec:dep}.
   \begin{figure}
   \centering
   \includegraphics[width=\hsize]{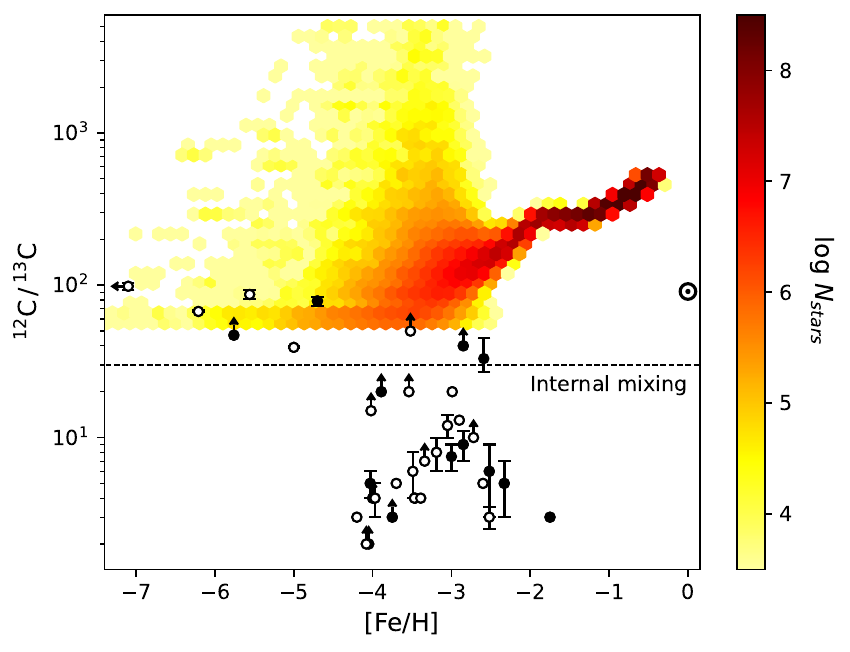}
      \caption{Same as Fig.~\ref{fig:SFR}, but for a model with H-He shell mergers in all 20 - 25 M$_\odot$ stars at \feh{\leq-3}. }
         \label{fig:SFR2}
   \end{figure}
\\In order to evaluate the impact of the H-He shell mergers on the chemical evolution, also in relation to the observations, we present in Fig.~\ref{fig:SFR2} the same chemical evolution model, but assuming that all stars in the range 20 - 25 M$_\odot$ at \feh{\leq-3} present a H-He shell merger. Since there are still not enough stellar models available, we used the yields for $^{12}$C and $^{13}$C from the zero-metal 25 M$_\odot$ stellar model with $v_\text{ini}=300\text{ km s}^{-1}$ from \cite{Roberti_2024}. The model in Fig.~\ref{fig:SFR2} predicts a much smaller \crat{} ratio at low metallicity, with most points between 50 - 300, much more in line with the observations that are all below \crat{<100}. These results show that a more frequent occurrence of shell mergers is preferable at low metallicity. This model still cannot reproduce the data below \crat{<30}, since we did not allow for extrapolation outside the stellar model predictions.

\subsection{Explanations for the low \crat{} ratio}\label{sec:dep}
\begin{figure}
   \centering
   \includegraphics[width=\hsize]{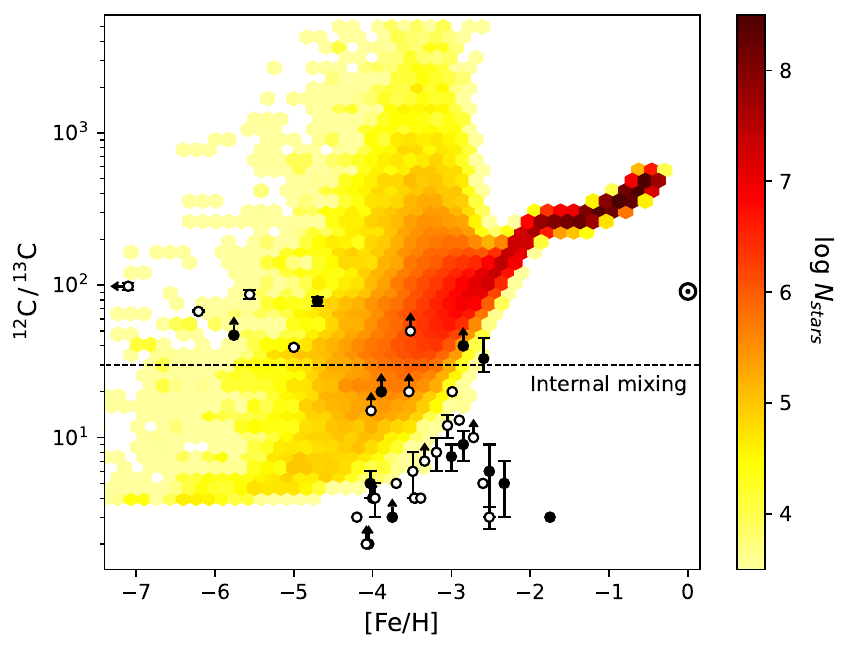}
      \caption{Same as Fig.~\ref{fig:SFR2}, but assuming that 20 - 25 M$_\odot$ stars at \feh{\leq-3} only eject the outer layers at the end of their lives. }
         \label{fig:SFR0Fe}
   \end{figure}

\begin{figure}
   \centering
   \includegraphics[width=\hsize]{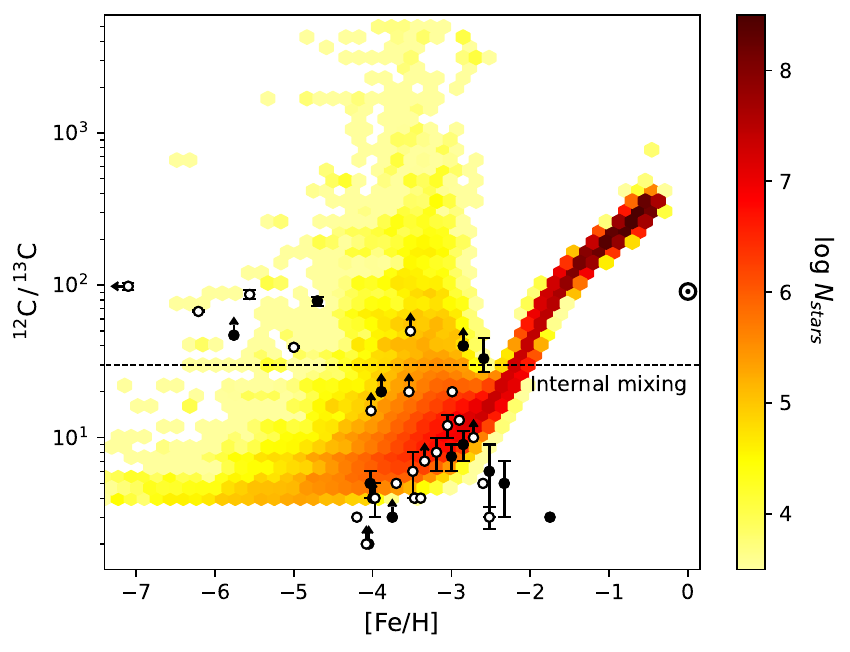}
      \caption{Same as Fig.~\ref{fig:SFR0Fe}, but assuming that H-He shell mergers in 20 - 25 M$_\odot$ stars at \feh{\leq-3} produce 10 times more $^{12}$C and $^{13}$C in the ejected outer layers than what currently predicted.}
         \label{fig:SFR10C}
   \end{figure}

We have seen that many observations of halo stars have a \crat{} ratio that lies below 30 \cite[see Fig.~\ref{fig:SFR} and][]{Molaro}. Stars in this region cannot be explained by the usual nucleosynthesis sites, therefore alternative physical mechanisms are required to explain their low \crat{}. Possible explanations involve modifications in the production sites, or in the surface abundances of the host stars, or a combination of both. \cite{Spite} have studied a sample of extremely metal poor RGB stars, finding evidence for modified surface abundances due to interior processes. A group of giants on the lower red giant branch, therefore less evolved (`unmixed giants'), showed a \crat{} ratio around 30 \cite[see also][]{2021A&A...652A..97S}, which therefore is sometimes described as the unmixed limit. 
Recently, stellar modelling has been investigating the possible depletion mechanisms in giant stars, through detailed comparison with individual observations. \cite{2023A&A...676A..19M}, using the \texttt{ATON} code, showed that 2.5 - 2.7 M$_\odot$ stars at solar metallicity display an important decrease in \crat{} surface abundance from about 90 to 20, only due to the first dredge-up during the RGB phase. Additionally, \cite{2024MNRAS.530..761M} and \cite{2024arXiv240805039N} found independently, the former employing the \texttt{ATON} and the latter the \texttt{PARSEC} code, that for stars $\leq1 \text{ M}_\odot$ with metallicity from solar down to $Z=10^{-4}$ (around \feh{=-2.6} in our galactic model), the initial \crat{=90} can be depleted to less than 10 due to the combined effects of thermohaline mixing and envelope overshooting. However, all these studies assume the solar ratio as the initial value, while at lower metallicity the initial surface ratio could be higher (see Fig.~\ref{fig:SFR}), therefore we cannot reconstruct the original surface abundance of giants in our range of interest.\\
In the sample we use here, 6 stars with \crat{<30} are not giants (see Section \ref{sec:obs} and Fig.~\ref{fig:SFR}). They should therefore retain their original surface \crat{}. 
A possible explanation for their low ratio is that the H-He shell mergers we explore in this study are able to produce much lower \crat{}. \cite{2016A&A...593A..36C} showed that an equilibrium value of $\sim4$ can be obtained in a one-zone nuclear model of CNO burning with mixing between H and He. \cite{2017A&A...605A..63C} implemented this in stellar models and were able to obtain yields of \crat{\sim 4} in massive-star outer layers by triggering a late H-He mixing event, at metallicity \feh{=-3.8} with very fast rotation ($v_\text{ini}=600\text{ - }700$ km s$^{-1}$). \cite{2021MNRAS.500.2685C} also found very low ratios \crat{\gtrsim1.5} in outer layers of zero-metal massive-star models without rotation. Considering the extra energy released by the proton ingestion, it is possible that the merging event produces enough energy to expel the outer layers before the SN explosion. However, a contribution by only outer layers, without deep layer ejection, would not be able to explain the observations. Iron is needed, from the inner layers of the same or other stars, to explain the [Fe/H] measurements. But if iron is ejected, so it is the abundant $^{12}$C in the inner layers, and \crat{} increases.\\
We show in Fig.~\ref{fig:SFR0Fe} the stochastic model when 20 - 25 M$_\odot$ stars at \feh{\leq-3}, the ones we assume produce shell mergers, only eject their outer regions. We used the yields from the same stellar models, but placed a mass cut at the base of the shell merger; this allowed to eject the equilibrium \crat{\sim4} in the shell, but no iron. Stars $<20$ M$_\odot$ still undergo full SN explosion and eject Fe and $^{12}$C. Therefore, stars at \feh{<-3} form from a mixture of progenitors with \crat{\sim4} but no iron or with large \crat{} and iron. The bulk of the model sits at \crat{>20} because the low \crat{} in the outer layer ejecta is mixed with the $^{12}$C-rich deep ejecta, increasing the ratio. A low probability region \crat{<20} is the result of volumes with many more progenitors 20 - 25 M$_\odot$ than 8 - 15 M$_\odot$, which is disfavoured but not unviable. As metallicity increases, more $^{12}$C accumulates in the ISM and the \crat{} ratio keeps increasing. These assumptions cannot reproduce the observations of our 6 dwarf stars at \crat{<15}, which remain completely outside the model. 
\\The only way of explaining all \crat{<30} with massive stars is to assume a larger production of $^{12}$C and $^{13}$C from H-He shell mergers, which can be obtained either increasing the yields of massive stars or their number. As an example of the possible impact, we show in Fig.~\ref{fig:SFR10C} the same chemical evolution model of Fig.~\ref{fig:SFR0Fe}, with shell mergers and outer layer ejection from 20 - 25 M$_\odot$ stars at \feh{\leq-3}, but assuming a production of $^{12}$C and $^{13}$C that is 10 times larger than what predicted by the models of \cite{Roberti_2024}. This is a tentative approach since such a large production is not based on any stellar model; however, this assumption pushes the \crat{} ratio to very low values that can be more in agreement with the observations. Similar results can be obtained assuming that more stars $\geq 20 \text{ M}_\odot$ can form compared to lower ones, therefore modifying the IMF. Evidences show that the IMF is possibly biased towards more massive stars at lower metallicity \citep{2018Sci...359...69S, 2018Natur.558..260Z}. \\
However, only changing the IMF would not be enough to have the larger production of $^{13}$C required to lower the \crat{} ratio. To show this, we recomputed some of the models presented above employing the IMF from \cite{2001MNRAS.322..231K}, which is more top-heavy compared to the one by \cite{1986FCPh...11....1S}; the models are shown in the appendix (Fig.~\ref{fig:Kroupa} and \ref{fig:Kroupa2}). The impact of the IMF is not significant: in particular, the bulk of the models, i.e.\ where the density of stars is larger, slightly shifts towards lower \crat{}, but the nucleosynthesis sources are the same as before and the IMF has not a strong enough effect on the number of massive stars to consistently lower \crat{}. Tests run with even more top-heavy IMF \citep{2018Sci...359...69S} produce the same results. 
\\Alternatively, it is possible to have a different threshold for the failed explosion or for the outer layer ejection. Indeed, the question of explodability for core-collapse supernovae in stars of different mass is still uncertain \citep{2021Natur.589...29B, 2023ApJ...949...17B}. Unfortunately, the grid of \cite{Roberti_2024} only encompasses two masses, so we cannot predict yet the effects of shell mergers in stars of different mass. In the grid of \cite{Limongi_2018} (see Fig.~\ref{fig:limyields}), there is a general trend of increasing \crat{} with larger mass, so a lower threshold for the failed explosion would decrease \crat{}. On the other hand, if higher-mass stars $> 25\text{ M}_\odot$ also produce shell mergers and only eject their outer layers, a wider mass range for outer layer ejection would be beneficial and lower \crat{}, since these stars could have an even larger production of $^{13}$C, given their mass, without ejecting the internal matter rich in $^{12}$C. Nevertheless, in order to shed more light on these issues additional stellar models of massive stars at low metallicity need to be produced.
\\Finally, we notice that assuming only outer layer ejection, the \crat{} remains low and cannot reproduce anymore the observations at \feh{<-4.5}, which have \crat{>30}. Shell mergers followed by complete supernova explosions were instead effective in explaining them, as we discussed in the previous Section. Therefore, it appears that both complete and incomplete ejection of material from stars with shell mergers are required to explain the entire range of \crat{} observed. \\
It is worth mentioning here that the assumption of outer layer ejection does not change significantly the chemical evolution of the other elements. CNO are still abundantly produced by lower mass stars and H-He shell mergers. Heavy elements are not produced in zero-metal stars in the first place, due to the lack of seed nuclei for neutron capture \citep[see][]{Roberti_2024}.

\subsection{Chemical evolution of the CNO elements}
\begin{figure}
   \centering
   \includegraphics[width=0.95\hsize]{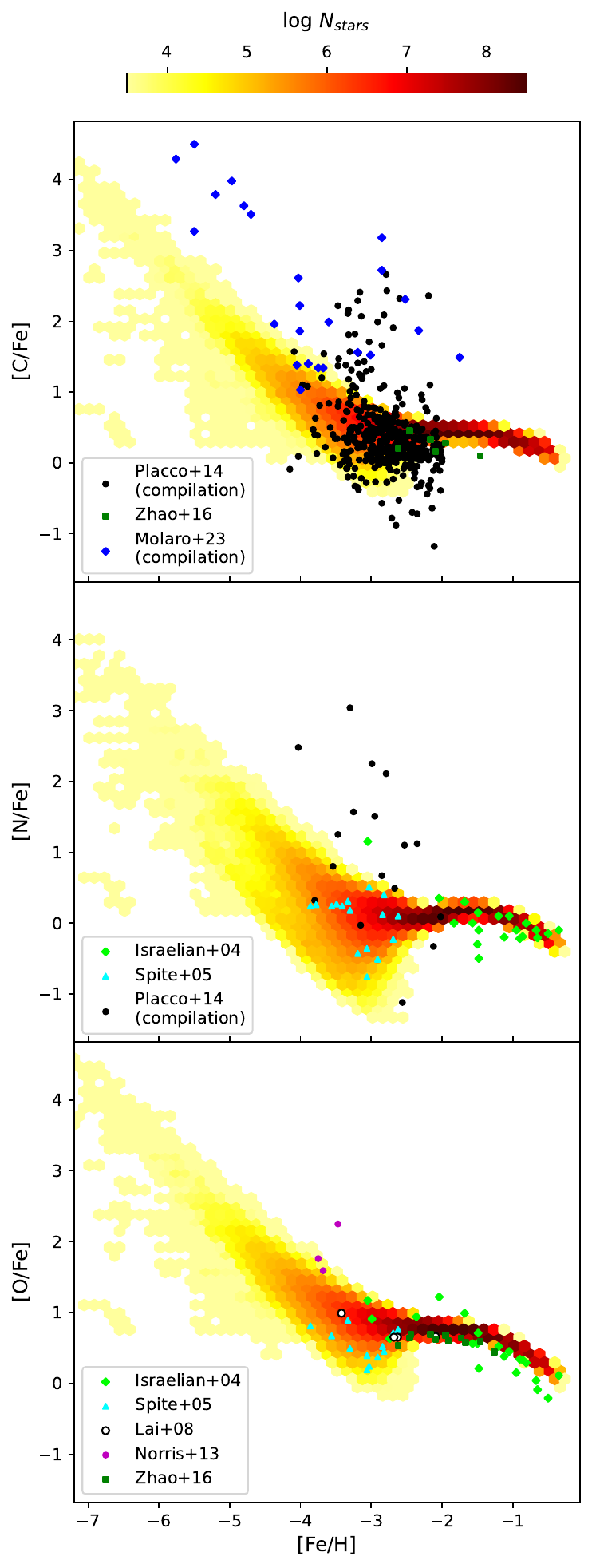}
      \caption{[C/Fe], [N/Fe], and [O/Fe] versus [Fe/H]. The stochastic model is the colour map, with the number of simulated stars in log scale. The dots are the observed dwarf or unmixed giant halo stars as presented in Section \ref{sec:obs}.}
         \label{fig:CNO}
   \end{figure}

\begin{figure}
   \centering
   \includegraphics[width=\hsize]{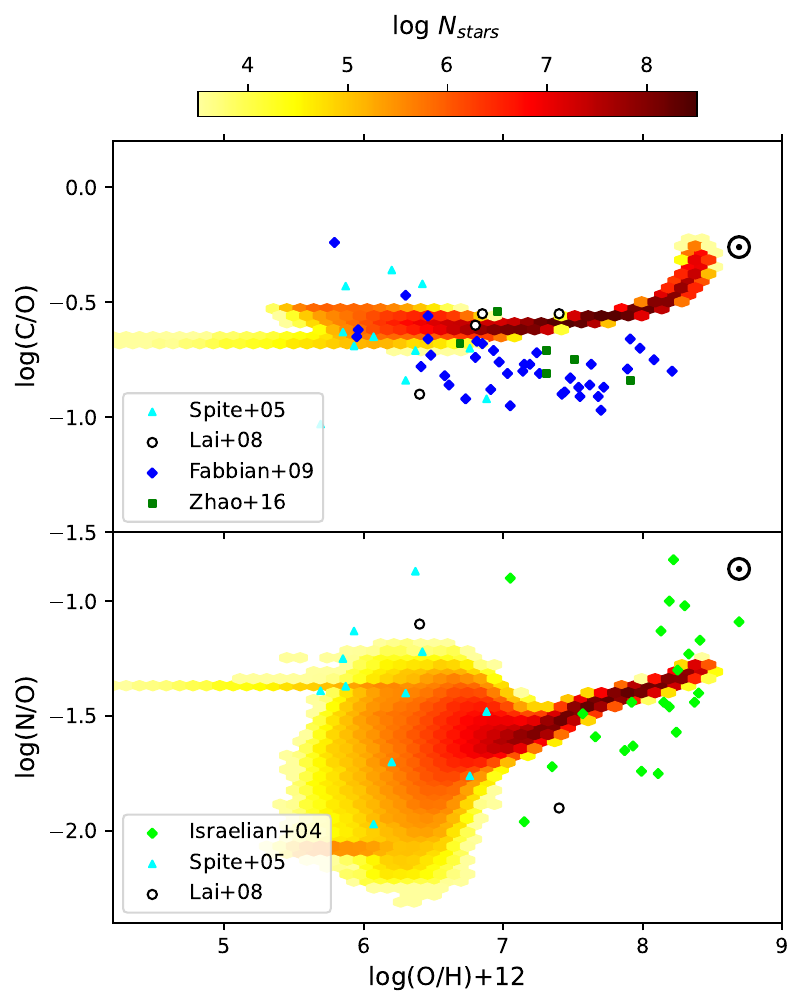}
      \caption{log(C/O) and log(N/O) versus log(O/H)+12. The stochastic model is the colour map, with the number of simulated stars in log scale. The dots are the observed dwarf or unmixed giant halo stars as presented in Section \ref{sec:obs}. The symbol of the Sun refers to the solar values \citep{2009ARA&A..47..481A}.}
         \label{fig:NOCO}
   \end{figure}

In Fig.~\ref{fig:CNO}, we present the predicted ratios [C/Fe], [N/Fe], and [O/Fe] for the model presented in Section~\ref{sec:13c}, Fig.~\ref{fig:SFR} together with the observations of dwarf stars or unmixed giants (see Section \ref{sec:obs}). These results are not significantly different from the other models with modified ejecta (see Section~\ref{sec:13c} and \ref{sec:dep}). Towards the lowest metallicity, the model predicts a long and narrow tail reaching very high values, due to the production from massive stars that eject mainly constant CNO and very little iron. Not many observations exist in this range except for carbon: the observed [C/Fe] always lie higher than the model for \feh{\lesssim -4}. This was also found by \cite{2016A&A...595A..91C} assuming the rotating yields of \cite{2002A&A...390..561M} and \cite{2007A&A...461..571H}. It could be due to the yields for massive stars underestimating C production, but also due to 1D LTE assumptions in the observations. In fact, 3D and NLTE corrections show a lower [C/Fe] measured at low metallicity \citep{2024arXiv240100697L}, down to $\sim1$ dex. For metallicity $-4<$ \feh{<-2}, the model is able to reproduce the normal stars, but not the CEMP stars ([C/Fe] > 1). The same result was found in the similar work of \cite{2010A&A...515A.102C}. Some of these stars have a very high C/H ratio, comparable to that of CEMP-s stars; they are sometimes called `super-CEMP' stars \citep{2014ApJ...791..116C}. Possible explanations for their high C involve binary interaction \citep{2014MNRAS.441.1217S, 2019A&A...621A.108A}, or an ISM that was not well mixed \citep{2003Natur.422..834B, 2003PASA...20..324C, 2010A&A...521A..30M}, or internal mixing \citep{2015A&A...580A..32M}.
\\Consequently, if on one hand zero-metal rotating massive stars with shell mergers can explain the \crat{} ratio, on the other hand, the [C/Fe] ratio is underestimated at the same metallicity. We interpret this result in the following way. In the stellar models of \cite{Roberti_2024}, the carbon yields generally increase with stellar rotation, both in the 15 and 25 M$_\odot$ models, up to around a factor 2. Therefore, the [C/Fe] observations can be explained assuming that primordial massive stars had even higher rotation velocity (\citealp{Roberti_2024} have run tests up to almost critical velocity), while the \crat{} observed in the same stars can be explained through shell merging events. The only way to combine both scenarios is to assume that H-He shell mergers are more common than what is currently found in stellar models, at least for extremely low metallicity. \cite{Roberti_2024} found that rotation seems to disfavour H-He shell mergers; however, recent findings from 3D stellar models \citep{2018MNRAS.481.2918M, Yadav_2020, 2024MNRAS.533..687R} seem to indicate that shell mergers may be more frequent once multi-dimensionality is taken into account.\\
For \feh{>-2}, the contribution from LIMS becomes important, so we do not extrapolate results from the model in this range. 
The picture is complicated by the fact that these elements are produced by both massive and low-mass stars, therefore a combination of the two is required to explain the observations in this metallicity range. This is beyond the scope of this work. \\
Figure \ref{fig:NOCO} shows the ratios in the form of log(C/O) and log(N/O) versus log(O/H)+12. In this way, the dependence from iron is removed and the elements are only studied in relation to each other. From the C/O ratio, we can see that the model is in line with the observations at lower metallicity, but is clearly overestimated at higher ones; this comes from the fact that the yields for carbon in set R by \cite{Limongi_2018} are generally overestimated compared to oxygen at \feh{>-2} (see Fig.~\ref{fig:CNO}), as also pointed out by \cite{2018MNRAS.476.3432P, 2019MNRAS.490.2838R, 2022IAUS..366...63K}.
From the C/O plot, we also notice that both model and observational trends are very narrow, because carbon and oxygen are produced in the same way. This is not the case for N/O, which instead presents a large spread at low metallicity both in the model and in the data. This can be explained considering that N has a primary production in rotating massive stars, being highly dependent on the stellar rotation, and the fact that massive stars above 25 M$_\odot$ contribute only through stellar wind, enriching the ISM with high N/O that is then diluted with material from stars < 25 M$_\odot$ \citep[see also][]{2010A&A...515A.102C}. The stochastic model can reproduce most of this spread. Going towards higher metallicity, the N/O trend follows the one of the observations, although not perfectly since we did not calibrate the model for these metallicities.

\subsection{Chemical evolution of Sr and Ba}
\begin{figure}
   \centering
   \includegraphics[width=0.95\hsize]{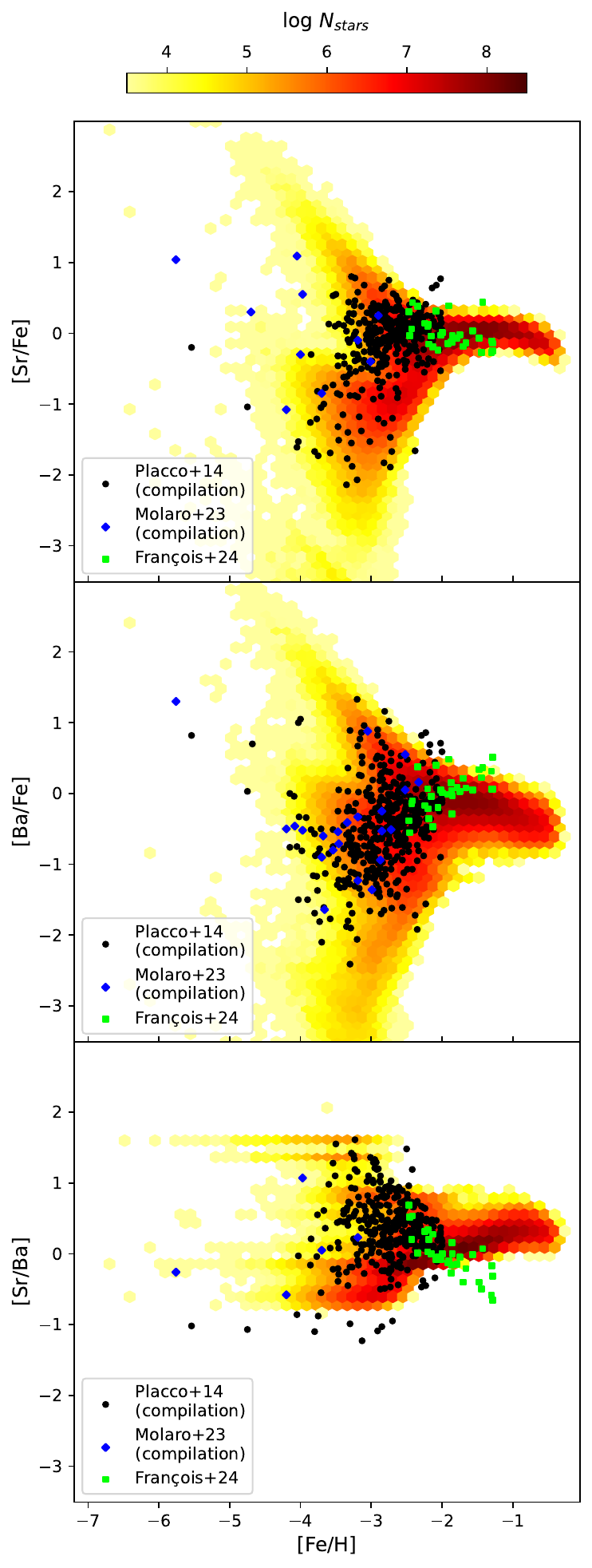}
      \caption{[Sr/Fe], [Ba/Fe], and [Sr/Ba] versus [Fe/H]. The stochastic model is the colour map, with the number of simulated stars in log scale. The dots are the observed halo stars as presented in Section \ref{sec:obs}.}
         \label{fig:SrBa}
   \end{figure}

In Fig.~\ref{fig:SrBa}, we show the predicted ratios [Sr/Fe], [Ba/Fe] and [Sr/Ba] from the same model compared to the observations. We remind that s-process material can be produced in rotating massive stars thanks to the rotation-induced mixing that stirs material between the H- and He-burning regions, enhancing the nucleosynthesis and neutron-capture thanks to the production of $^{14}$N and its conversion to $^{22}$Ne as neutron source \citep[see][]{Limongi_2018}. These effects are even more important at low metallicity, where stars are more compact and tend to rotate faster, and where the neutron-to-seed ratio drastically increases, favouring the production of heavier nuclei \citep[see the discussion on s-process in][]{1998ApJ...497..388G}. This is of crucial importance, as massive stars are thought to only contribute to the weak s-process component (Sr-Y-Zr). In this way, they can also contribute to the main component (Ba-La-Ce), but not to the heavy one (Pb). Although zero-metal stars cannot synthesize s-process elements due to their lack of iron seeds for neutron capture, already the next generation of stars (\feh{=-5} in our model) begins the production \citep[see][]{Roberti_2024}.\\
The stochastic model is able to reproduce the majority of the observations, even at extremely low metallicity. The long horizontal lines in [Sr/Ba] towards low metallicity come from the fact that few masses are present in the grid of yields below \feh{<-3}.
\\It is meaningful here to make a comparison between these results and the ones presented in \cite{Rizzuti2021} for the same abundance ratios. The model of \cite{Rizzuti2021} is able to better reproduce the trend and spread of Sr and Ba in the observations, thanks to the fact that they calibrated the massive-star rotation velocity, while here we use a simplified distribution. Above \feh{>-2}, the model is less accurate, because it has not been calibrated for this metallicity range, and the impact from AGB stars begins to be important. The new data from \cite{2024A&A...686A.295F} show an important increase in Ba at higher metallicity compared to the model predictions: this is an interesting point that will require additional investigation in the future.

\section{Conclusions}
In this paper, we investigated the production of light elements C, N, O, and the \crat{} ratio in zero-metal and extremely low metallicity stars. This has been required by the recent observation campaigns of stars in the Galactic halo. Explaining these \crat{} ratios at low metallicity is challenging, considering that the measurements span the range $3<$ \crat{<100}, requiring a consistent production of $^{13}$C that must be primary. \\
This production of $^{13}$C is expected to come from massive stars, which are the main contributors at these low metallicities. Over the years, stellar modelling studies have been showing that $^{13}$C can be produced in a primary fashion by the interaction between H and He shells, assuming a mixing between these layers that can be favoured by rotation \citep{2007A&A...461..571H,2017A&A...605A..63C, Limongi_2018, 2024A&A...687A.307T} or convection \citep{2012ApJS..199...38L,2021MNRAS.500.2685C}, or a combination of both \citep{Roberti_2024}. However, it is presently believed that stars at low metallicity need to be fast rotators, for both physical \citep{2002A&A...390..561M} and nucleosynthesis \citep{2006A&A...449L..27C, 2013A&A...553A..51C} reasons.\\
In this study, we have implemented in the \texttt{GEMS} chemical evolution model, the latest version of the stochastic model by \cite{2010A&A...515A.102C}, the most up-to-date nucleosynthesis sources available in the literature, including the recent work by \cite{Roberti_2024}, which presents stellar evolution models for zero-metal and low-metallicity rotating massive stars. We find no need to include the contribution from very massive (>$100\text{ M}_\odot$) and supermassive (>$1000\text{ M}_\odot$) Pop III stars, and in general we treat Pop III stars as if they have no qualitative difference from low-metallicity stars, except the different metal content. We show that thanks to the H-He shell merging event occurring in some massive stars, the chemical evolution model is able to reproduce the observations at \crat{>30}. The model is in best agreement with the observations if we assume that H-He shell mergers occur in all 20 - 25 M$_\odot$ stars at \feh{\leq-3}. The evolution of [CNO/Fe] can also be adequately reproduced for low-metallicity stars except for CEMP and super-CEMP stars. The discrepancy with the observations at the lowest metallicity in [C/Fe] and C/O can be attributed to 3D and NLTE effects in the measurements \citep{2024arXiv240100697L}. Additionally, also the evolution of heavy elements Sr and Ba, produced in part by the same sources but not calibrated in this work, is reproduced.\\
Explaining the observations below \crat{<30} still remains an open question. Different stellar modelling studies have been showing that a very low \crat{} ratio can be obtained in stellar winds or outer layers of massive stars \cite[e.g.][]{2007A&A...461..571H, 2017A&A...605A..63C, 2021MNRAS.500.2685C}. The ejection of only the outer layers can be caused by the energy produced at the base of the shell merger \citep[see][]{2021MNRAS.500.2685C}, followed by a collapse of the star with no SN explosion. However, to reproduce at the same time both the \crat{} and [Fe/H] measured in CEMP stars, it is necessary to assume that they formed from both inner and outer layers of SN ejections, from the same or different stars. \\
We have shown here that observations at \crat{<30} can be explained assuming that the $^{13}$C-rich outer layers are expelled from the star with no subsequent SN explosion, and then mixed with Fe from other SN remnants, but only if the production of $^{13}$C is 10 times larger than what currently predicted, which is at best $\sim 10^{-2}\text{ M}_\odot$ in 20 - 25 M$_\odot$ shell-merger stars. This cannot be achieved only assuming that the number of massive stars is larger than what is currently assumed, although observational evidence pointed out that the IMF could be biased towards more massive stars at lower metallicity \citep{2018Sci...359...69S, 2018Natur.558..260Z, 2021MNRAS.503.6026R}. Another possibility is that also stars $> 25\text{ M}_\odot$ produce H-He shell mergers and eject them before their final collapse, given that the energy released from the merging is enough to eject the outer layers. Although these stars are fewer in number, their higher mass could contribute to a larger production of $^{13}$C.\\ 
However, to keep reproducing also the \crat{>30} observations at lowest metallicity it is necessary to assume that some stars also formed from complete SN ejecta. This is possible either if it is present a threshold around the mass range 20 - 25 M$_\odot$ that separates complete and incomplete ejection, or if the ejection occurs differently in stars with the same mass, depending on physical properties such as rotation or magnetic fields \citep[see][]{2023MNRAS.518.3622V, 2024Univ...10..148B}. Therefore, we make here a distinction between faint and failed supernovae: the former still eject small quantities of iron, necessary to explain the [Fe/H] measured in CEMP stars; the latter need to eject only the outer layers, enriching the ISM of low \crat{} material. \\
Overall, the work we presented here shows the need for H-He shell mergers to explain the nucleosynthesis of the first stars. The occurrence of H-He interacting shells has been reported for a long time in stellar modelling, both in zero-metal \citep{1982ASIC...90...79W, 2010ApJ...724..341H, 2012ApJS..199...38L, 2021MNRAS.500.2685C} and low-metallicity \citep{2007A&A...461..571H, 2017A&A...605A..63C, 2018MNRAS.480..538R} stars. It is important to underline that they are expected to be more frequent at lower metallicity, due to the lower opacity of the star, lower entropy jump and smaller distance between the shells both in radius and in mass. \cite{Roberti_2024} concluded that stellar rotation seems to disfavour the occurrence of H-He shell mergers. However, these occurrences are strongly dependent on the convective and overshoot prescriptions used in the stellar models. \\
Recent findings from both 1D and 3D stellar models seem to indicate that the mixing-length-theory \citep{1958ZA.....46..108B} normally used to reproduce convection may be inaccurate for advanced phases of massive stars \citep{2018arXiv181004659A, 2022Rizzuti, 2024MNRAS.531.4293G} and H-He interacting environments \citep{2011ApJ...727...89H, Herwig_2014, 2016MNRAS.455.3848J}. Convective-boundary-mixing (overshoot) prescriptions seem equally underestimated \citep{2021Scott,2019Cristini,2023Rizzuti}. While 3D simulations of H-ingestion environments have only been focusing on AGB stars \citep{2011A&A...533A..53M, 2011ApJ...742..121S, Herwig_2014}, late-type C-Ne-O shell mergers have already been confirmed to occur also in 3D stellar models \citep{Yadav_2020, 2024MNRAS.533..687R}. The implementation of stronger convection in the stellar models would naturally result in more frequent occurrence of shell mergers.\\
It is clear that to shed more light on these issues, additional stellar models of zero-metal and low-metallicity rotating massive stars are necessary. This is especially important for modelling the Galactic chemical evolution, which is largely based on detailed grids of the stellar yields. Even though there are still many uncertainties about the parameters to be used in stellar models, results from Galactic archaeology can provide useful constraints on the physics of stars, such as their mass distribution, rotation velocity, frequency of ingestion events, and fate. Therefore, having better constraints on the nucleosynthesis of the first stars will also have important implications for our knowledge of their physics, which today is still largely obscure. 


\begin{acknowledgements}
FR, GC, and LM acknowledge financial support under the National Recovery and Resilience Plan (NRRP), Mission 4, Component 2, Investment 1.1, Call for tender No. 104 published on 2.2.2022 by the Italian Ministry of University and Research (MUR), funded by the European Union - NextGenerationEU - Project ‘Cosmic POT’ (PI: L. Magrini) Grant Assignment Decree No. 2022X4TM3H  by the Italian Ministry of Ministry of University and Research (MUR). GC and LM  thank INAF for the support (Large Grant 2023 EPOCH) and for the MiniGrant 2022 Checs. LR acknowledges the support from the NKFI via K-project 138031 and the Lendület Program LP2023-10 of the Hungarian Academy of Sciences. This work has been partially supported by the Italian grants “Premiale 2015 FIGARO” (PI: Gianluca Gemme). We acknowledge support from PRIN MUR 2022 (20224MNC5A), "Life, death and after-death of massive stars", funded by European Union - Next Generation EU. FM thanks INAF for the 1.05.12.06.05 Theory Grant - Galactic archaeology with radioactive and stable nuclei. FM acknowledges also support from Project PRIN MUR 2022 (code 2022ARWP9C) “Early Formation and Evolution of Bulge and HalO (EFEBHO)” (PI: M. Marconi). This work was also partially supported by the European Union (ChETEC-INFRA, project no. 101008324).\end{acknowledgements}

%
   \bibliographystyle{aa} 
   \bibliography{aa.bib} 

\begin{thebibliography}{167}
\expandafter\ifx\csname natexlab\endcsname\relax\def\natexlab#1{#1}\fi

\bibitem[{{Aguado} {et~al.}(2023){Aguado}, {Caffau}, {Molaro}, {Allende
  Prieto}, {Bonifacio}, {Gonz{\'a}lez Hern{\'a}ndez}, {Rebolo}, {Salvadori},
  {Zapatero Osorio}, {Cristiani}, {Pepe}, {Santos}, {Cupani}, {Di Marcantonio},
  {D'Odorico}, {Lovis}, {Nunes}, {Martins}, {Milakovi}, {Rodrigues}, {Schmidt},
  {Sozzetti}, \& {Su{\'a}rez Mascare{\~n}o}}]{2023A&A...669L...4A}
{Aguado}, D.~S., {Caffau}, E., {Molaro}, P., {et~al.} 2023, \aap, 669, L4

\bibitem[{{Aguado} {et~al.}(2018){Aguado}, {Gonz{\'a}lez Hern{\'a}ndez},
  {Allende Prieto}, \& {Rebolo}}]{2018ApJ...852L..20A}
{Aguado}, D.~S., {Gonz{\'a}lez Hern{\'a}ndez}, J.~I., {Allende Prieto}, C., \&
  {Rebolo}, R. 2018, \apjl, 852, L20

\bibitem[{{Aguado} {et~al.}(2019){Aguado}, {Gonz{\'a}lez Hern{\'a}ndez},
  {Allende Prieto}, \& {Rebolo}}]{2019ApJ...874L..21A}
{Aguado}, D.~S., {Gonz{\'a}lez Hern{\'a}ndez}, J.~I., {Allende Prieto}, C., \&
  {Rebolo}, R. 2019, \apjl, 874, L21

\bibitem[{{Aguado} {et~al.}(2022){Aguado}, {Molaro}, {Caffau}, {Gonz{\'a}lez
  Hern{\'a}ndez}, {Zapatero Osorio}, {Bonifacio}, {Allende Prieto}, {Rebolo},
  {Damasso}, {Su{\'a}rez Mascare{\~n}o}, {Howell}, {Furlan}, {Cristiani},
  {Cupani}, {Di Marcantonio}, {D'Odorico}, {Lovis}, {Martins}, {Milakovi},
  {Murphy}, {Nunes}, {Pepe}, {Santos}, {Schmidt}, \&
  {Sozzetti}}]{2022A&A...668A..86A}
{Aguado}, D.~S., {Molaro}, P., {Caffau}, E., {et~al.} 2022, \aap, 668, A86

\bibitem[{{Allen} {et~al.}(2012){Allen}, {Ryan}, {Rossi}, {Beers}, \&
  {Tsangarides}}]{2012A&A...548A..34A}
{Allen}, D.~M., {Ryan}, S.~G., {Rossi}, S., {Beers}, T.~C., \& {Tsangarides},
  S.~A. 2012, \aap, 548, A34

\bibitem[{{Aoki} {et~al.}(2002){Aoki}, {Norris}, {Ryan}, {Beers}, \&
  {Ando}}]{2002ApJ...567.1166A}
{Aoki}, W., {Norris}, J.~E., {Ryan}, S.~G., {Beers}, T.~C., \& {Ando}, H. 2002,
  \apj, 567, 1166

\bibitem[{{Arentsen} {et~al.}(2019){Arentsen}, {Starkenburg}, {Shetrone},
  {Venn}, {Depagne}, \& {McConnachie}}]{2019A&A...621A.108A}
{Arentsen}, A., {Starkenburg}, E., {Shetrone}, M.~D., {et~al.} 2019, \aap, 621,
  A108

\bibitem[{{Argast} {et~al.}(2002){Argast}, {Samland}, {Thielemann}, \&
  {Gerhard}}]{2002A&A...388..842A}
{Argast}, D., {Samland}, M., {Thielemann}, F.~K., \& {Gerhard}, O.~E. 2002,
  \aap, 388, 842

\bibitem[{{Argast} {et~al.}(2004){Argast}, {Samland}, {Thielemann}, \&
  {Qian}}]{2004A&A...416..997A}
{Argast}, D., {Samland}, M., {Thielemann}, F.~K., \& {Qian}, Y.~Z. 2004, \aap,
  416, 997

\bibitem[{Arnett {et~al.}(2018)Arnett, Hirschi, Campbell, Mocák, Georgy,
  Meakin, Cristini, Scott, Kaiser, \& Viallet}]{2018arXiv181004659A}
Arnett, W.~D., Hirschi, R., Campbell, S.~W., {et~al.} 2018, preprint

\bibitem[{{Asplund} {et~al.}(2009){Asplund}, {Grevesse}, {Sauval}, \&
  {Scott}}]{2009ARA&A..47..481A}
{Asplund}, M., {Grevesse}, N., {Sauval}, A.~J., \& {Scott}, P. 2009, \araa, 47,
  481

\bibitem[{{Ayres} {et~al.}(2013){Ayres}, {Lyons}, {Ludwig}, {Caffau}, \&
  {Wedemeyer-B{\"o}hm}}]{2013ApJ...765...46A}
{Ayres}, T.~R., {Lyons}, J.~R., {Ludwig}, H.~G., {Caffau}, E., \&
  {Wedemeyer-B{\"o}hm}, S. 2013, \apj, 765, 46

\bibitem[{{Beers} {et~al.}(1985){Beers}, {Preston}, \&
  {Shectman}}]{1985AJ.....90.2089B}
{Beers}, T.~C., {Preston}, G.~W., \& {Shectman}, S.~A. 1985, \aj, 90, 2089

\bibitem[{{Beers} {et~al.}(1992){Beers}, {Preston}, \&
  {Shectman}}]{1992AJ....103.1987B}
{Beers}, T.~C., {Preston}, G.~W., \& {Shectman}, S.~A. 1992, \aj, 103, 1987

\bibitem[{{Bessell} \& {Norris}(1984)}]{1984ApJ...285..622B}
{Bessell}, M.~S. \& {Norris}, J. 1984, \apj, 285, 622

\bibitem[{{Boccioli} \& {Roberti}(2024)}]{2024Univ...10..148B}
{Boccioli}, L. \& {Roberti}, L. 2024, Universe, 10, 148

\bibitem[{{Boccioli} {et~al.}(2023){Boccioli}, {Roberti}, {Limongi}, {Mathews},
  \& {Chieffi}}]{2023ApJ...949...17B}
{Boccioli}, L., {Roberti}, L., {Limongi}, M., {Mathews}, G.~J., \& {Chieffi},
  A. 2023, \apj, 949, 17

\bibitem[{{B{\"o}hm-Vitense}(1958)}]{1958ZA.....46..108B}
{B{\"o}hm-Vitense}, E. 1958, \zap, 46, 108

\bibitem[{{Bonifacio} {et~al.}(2015){Bonifacio}, {Caffau}, {Spite}, {Limongi},
  {Chieffi}, {Klessen}, {Fran{\c{c}}ois}, {Molaro}, {Ludwig}, {Zaggia},
  {Spite}, {Plez}, {Cayrel}, {Christlieb}, {Clark}, {Glover}, {Hammer}, {Koch},
  {Monaco}, {Sbordone}, \& {Steffen}}]{2015A&A...579A..28B}
{Bonifacio}, P., {Caffau}, E., {Spite}, M., {et~al.} 2015, \aap, 579, A28

\bibitem[{{Bonifacio} {et~al.}(2018){Bonifacio}, {Caffau}, {Spite}, {Spite},
  {Sbordone}, {Monaco}, {Fran{\c{c}}ois}, {Plez}, {Molaro}, {Gallagher},
  {Cayrel}, {Christlieb}, {Klessen}, {Koch}, {Ludwig}, {Steffen}, {Zaggia}, \&
  {Abate}}]{2018A&A...612A..65B}
{Bonifacio}, P., {Caffau}, E., {Spite}, M., {et~al.} 2018, \aap, 612, A65

\bibitem[{{Bonifacio} {et~al.}(2003){Bonifacio}, {Limongi}, \&
  {Chieffi}}]{2003Natur.422..834B}
{Bonifacio}, P., {Limongi}, M., \& {Chieffi}, A. 2003, \nat, 422, 834

\bibitem[{{Bonifacio} {et~al.}(1998){Bonifacio}, {Molaro}, {Beers}, \&
  {Vladilo}}]{1998A&A...332..672B}
{Bonifacio}, P., {Molaro}, P., {Beers}, T.~C., \& {Vladilo}, G. 1998, \aap,
  332, 672

\bibitem[{{Bressan} {et~al.}(2012){Bressan}, {Marigo}, {Girardi}, {Salasnich},
  {Dal Cero}, {Rubele}, \& {Nanni}}]{2012MNRAS.427..127B}
{Bressan}, A., {Marigo}, P., {Girardi}, L., {et~al.} 2012, \mnras, 427, 127

\bibitem[{{Burrows} \& {Vartanyan}(2021)}]{2021Natur.589...29B}
{Burrows}, A. \& {Vartanyan}, D. 2021, \nat, 589, 29

\bibitem[{{Busso} {et~al.}(1999){Busso}, {Gallino}, \&
  {Wasserburg}}]{1999ARA&A..37..239B}
{Busso}, M., {Gallino}, R., \& {Wasserburg}, G.~J. 1999, \araa, 37, 239

\bibitem[{{Caffau} {et~al.}(2013){Caffau}, {Bonifacio}, {Sbordone},
  {Fran{\c{c}}ois}, {Monaco}, {Spite}, {Plez}, {Cayrel}, {Christlieb}, {Clark},
  {Glover}, {Klessen}, {Koch}, {Ludwig}, {Spite}, {Steffen}, \&
  {Zaggia}}]{2013A&A...560A..71C}
{Caffau}, E., {Bonifacio}, P., {Sbordone}, L., {et~al.} 2013, \aap, 560, A71

\bibitem[{{Caffau} {et~al.}(2016){Caffau}, {Bonifacio}, {Spite}, {Spite},
  {Monaco}, {Sbordone}, {Fran{\c{c}}ois}, {Gallagher}, {Plez}, {Zaggia},
  {Ludwig}, {Cayrel}, {Koch}, {Steffen}, {Salvadori}, {Klessen}, {Glover}, \&
  {Christlieb}}]{2016A&A...595L...6C}
{Caffau}, E., {Bonifacio}, P., {Spite}, M., {et~al.} 2016, \aap, 595, L6

\bibitem[{{Cavallo} {et~al.}(2021){Cavallo}, {Cescutti}, \&
  {Matteucci}}]{2021MNRAS.503....1C}
{Cavallo}, L., {Cescutti}, G., \& {Matteucci}, F. 2021, \mnras, 503, 1

\bibitem[{{Cescutti}(2008)}]{2008A&A...481..691C}
{Cescutti}, G. 2008, \aap, 481, 691

\bibitem[{{Cescutti} {et~al.}(2022){Cescutti}, {Bonifacio}, {Caffau}, {Monaco},
  {Franchini}, {Lombardo}, {Matas Pinto}, {Lucertini}, {François}, {Spitoni,
  E.}, {Lallement, R.}, {Sbordone, L.}, {Mucciarelli, A.}, {Spite, M.},
  {Hansen, C. J.}, {Di Marcantonio, P.}, {Kučinskas, A.}, {Dobrovolskas, V.},
  {Korn, A. J.}, {Valentini, M.}, {Magrini, L.}, {Cristallo, S.}, \&
  {Matteucci, F.}}]{CescuttiMince}
{Cescutti}, G., {Bonifacio}, P., {Caffau}, E., {et~al.} 2022, \aap, 668, A168

\bibitem[{{Cescutti} \& {Chiappini}(2010)}]{2010A&A...515A.102C}
{Cescutti}, G. \& {Chiappini}, C. 2010, \aap, 515, A102

\bibitem[{{Cescutti} \& {Chiappini}(2014)}]{2014A&A...565A..51C}
{Cescutti}, G. \& {Chiappini}, C. 2014, \aap, 565, A51

\bibitem[{{Cescutti} {et~al.}(2013){Cescutti}, {Chiappini}, {Hirschi},
  {Meynet}, \& {Frischknecht}}]{2013A&A...553A..51C}
{Cescutti}, G., {Chiappini}, C., {Hirschi}, R., {Meynet}, G., \&
  {Frischknecht}, U. 2013, \aap, 553, A51

\bibitem[{{Cescutti} {et~al.}(2015){Cescutti}, {Romano}, {Matteucci},
  {Chiappini}, \& {Hirschi}}]{2015A&A...577A.139C}
{Cescutti}, G., {Romano}, D., {Matteucci}, F., {Chiappini}, C., \& {Hirschi},
  R. 2015, \aap, 577, A139

\bibitem[{{Cescutti} {et~al.}(2016){Cescutti}, {Valentini}, {Fran{\c{c}}ois},
  {Chiappini}, {Depagne}, {Christlieb}, \& {Cort{\'e}s}}]{2016A&A...595A..91C}
{Cescutti}, G., {Valentini}, M., {Fran{\c{c}}ois}, P., {et~al.} 2016, \aap,
  595, A91

\bibitem[{{Chamberlain} \& {Aller}(1951)}]{1951ApJ...114...52C}
{Chamberlain}, J.~W. \& {Aller}, L.~H. 1951, \apj, 114, 52

\bibitem[{{Chen} {et~al.}(2014){Chen}, {Girardi}, {Bressan}, {Marigo},
  {Barbieri}, \& {Kong}}]{2014MNRAS.444.2525C}
{Chen}, Y., {Girardi}, L., {Bressan}, A., {et~al.} 2014, \mnras, 444, 2525

\bibitem[{{Chiappini} {et~al.}(2008){Chiappini}, {Ekstr{\"o}m}, {Meynet},
  {Hirschi}, {Maeder}, \& {Charbonnel}}]{2008A&A...479L...9C}
{Chiappini}, C., {Ekstr{\"o}m}, S., {Meynet}, G., {et~al.} 2008, \aap, 479, L9

\bibitem[{{Chiappini} {et~al.}(2006){Chiappini}, {Hirschi}, {Meynet},
  {Ekstr{\"o}m}, {Maeder}, \& {Matteucci}}]{2006A&A...449L..27C}
{Chiappini}, C., {Hirschi}, R., {Meynet}, G., {et~al.} 2006, \aap, 449, L27

\bibitem[{{Chieffi} \& {Limongi}(2003)}]{2003PASA...20..324C}
{Chieffi}, A. \& {Limongi}, M. 2003, \pasa, 20, 324

\bibitem[{{Chieffi} \& {Limongi}(2013)}]{2013ApJ...764...21C}
{Chieffi}, A. \& {Limongi}, M. 2013, \apj, 764, 21

\bibitem[{{Chieffi} \& {Limongi}(2020)}]{2020ApJ...890...43C}
{Chieffi}, A. \& {Limongi}, M. 2020, \apj, 890, 43

\bibitem[{{Choplin} {et~al.}(2017){Choplin}, {Ekstr{\"o}m}, {Meynet}, {Maeder},
  {Georgy}, \& {Hirschi}}]{2017A&A...605A..63C}
{Choplin}, A., {Ekstr{\"o}m}, S., {Meynet}, G., {et~al.} 2017, \aap, 605, A63

\bibitem[{{Choplin} {et~al.}(2016){Choplin}, {Maeder}, {Meynet}, \&
  {Chiappini}}]{2016A&A...593A..36C}
{Choplin}, A., {Maeder}, A., {Meynet}, G., \& {Chiappini}, C. 2016, \aap, 593,
  A36

\bibitem[{{Clarkson} \& {Herwig}(2021)}]{2021MNRAS.500.2685C}
{Clarkson}, O. \& {Herwig}, F. 2021, \mnras, 500, 2685

\bibitem[{{Cohen} {et~al.}(2004){Cohen}, {Christlieb}, {McWilliam}, {Shectman},
  {Thompson}, {Wasserburg}, {Ivans}, {Dehn}, {Karlsson}, \&
  {Melendez}}]{2004ApJ...612.1107C}
{Cohen}, J.~G., {Christlieb}, N., {McWilliam}, A., {et~al.} 2004, \apj, 612,
  1107

\bibitem[{{Collins} {et~al.}(2018){Collins}, {M{\"u}ller}, \&
  {Heger}}]{2018MNRAS.473.1695C}
{Collins}, C., {M{\"u}ller}, B., \& {Heger}, A. 2018, \mnras, 473, 1695

\bibitem[{{Cooke} \& {Madau}(2014)}]{2014ApJ...791..116C}
{Cooke}, R.~J. \& {Madau}, P. 2014, \apj, 791, 116

\bibitem[{{Cristallo} {et~al.}(2011){Cristallo}, {Piersanti}, {Straniero},
  {Gallino}, {Dom{\'\i}nguez}, {Abia}, {Di Rico}, {Quintini}, \&
  {Bisterzo}}]{2011ApJS..197...17C}
{Cristallo}, S., {Piersanti}, L., {Straniero}, O., {et~al.} 2011, \apjs, 197,
  17

\bibitem[{{Cristallo} {et~al.}(2009){Cristallo}, {Straniero}, {Gallino},
  {Piersanti}, {Dom{\'\i}nguez}, \& {Lederer}}]{2009ApJ...696..797C}
{Cristallo}, S., {Straniero}, O., {Gallino}, R., {et~al.} 2009, \apj, 696, 797

\bibitem[{{Cristallo} {et~al.}(2015){Cristallo}, {Straniero}, {Piersanti}, \&
  {Gobrecht}}]{2015ApJS..219...40C}
{Cristallo}, S., {Straniero}, O., {Piersanti}, L., \& {Gobrecht}, D. 2015,
  \apjs, 219, 40

\bibitem[{Cristini {et~al.}(2019)Cristini, Hirschi, Meakin, Arnett, Georgy, \&
  Walkington}]{2019Cristini}
Cristini, A., Hirschi, R., Meakin, C., {et~al.} 2019, \mnras, 484, 4645

\bibitem[{{Curtis} {et~al.}(2019){Curtis}, {Ebinger}, {Fr{\"o}hlich}, {Hempel},
  {Perego}, {Liebend{\"o}rfer}, \& {Thielemann}}]{2019ApJ...870....2C}
{Curtis}, S., {Ebinger}, K., {Fr{\"o}hlich}, C., {et~al.} 2019, \apj, 870, 2

\bibitem[{{Deng} {et~al.}(2012){Deng}, {Newberg}, {Liu}, {Carlin}, {Beers},
  {Chen}, {Chen}, {Christlieb}, {Grillmair}, {Guhathakurta}, {Han}, {Hou},
  {Lee}, {L{\'e}pine}, {Li}, {Liu}, {Pan}, {Sellwood}, {Wang}, {Wang}, {Yang},
  {Yanny}, {Zhang}, {Zhang}, {Zheng}, \& {Zhu}}]{2012RAA....12..735D}
{Deng}, L.-C., {Newberg}, H.~J., {Liu}, C., {et~al.} 2012, Research in
  Astronomy and Astrophysics, 12, 735

\bibitem[{{Ekstr{\"o}m} {et~al.}(2008){Ekstr{\"o}m}, {Meynet}, {Chiappini},
  {Hirschi}, \& {Maeder}}]{2008A&A...489..685E}
{Ekstr{\"o}m}, S., {Meynet}, G., {Chiappini}, C., {Hirschi}, R., \& {Maeder},
  A. 2008, \aap, 489, 685

\bibitem[{{Fabbian} {et~al.}(2009){Fabbian}, {Nissen}, {Asplund}, {Pettini}, \&
  {Akerman}}]{2009A&A...500.1143F}
{Fabbian}, D., {Nissen}, P.~E., {Asplund}, M., {Pettini}, M., \& {Akerman}, C.
  2009, \aap, 500, 1143

\bibitem[{{Fran{\c{c}}ois} {et~al.}(2024){Fran{\c{c}}ois}, {Cescutti},
  {Bonifacio}, {Caffau}, {Monaco}, {Steffen}, {Puschnig}, {Calura},
  {Cristallo}, {Di Marcantonio}, {Dobrovolskas}, {Franchini}, {Gallagher},
  {Hansen}, {Korn}, {Ku{\v{c}}inskas}, {Lallement}, {Lombardo}, {Lucertini},
  {Magrini}, {Matas Pinto}, {Matteucci}, {Mucciarelli}, {Sbordone}, {Spite},
  {Spitoni}, \& {Valentini}}]{2024A&A...686A.295F}
{Fran{\c{c}}ois}, P., {Cescutti}, G., {Bonifacio}, P., {et~al.} 2024, \aap,
  686, A295

\bibitem[{{Frebel} {et~al.}(2015){Frebel}, {Chiti}, {Ji}, {Jacobson}, \&
  {Placco}}]{2015ApJ...810L..27F}
{Frebel}, A., {Chiti}, A., {Ji}, A.~P., {Jacobson}, H.~R., \& {Placco}, V.~M.
  2015, \apjl, 810, L27

\bibitem[{{Frebel} {et~al.}(2019){Frebel}, {Ji}, {Ezzeddine}, {Hansen},
  {Chiti}, {Thompson}, \& {Merle}}]{2019ApJ...871..146F}
{Frebel}, A., {Ji}, A.~P., {Ezzeddine}, R., {et~al.} 2019, \apj, 871, 146

\bibitem[{{Frischknecht} {et~al.}(2016){Frischknecht}, {Hirschi}, {Pignatari},
  {Maeder}, {Meynet}, {Chiappini}, {Thielemann}, {Rauscher}, {Georgy}, \&
  {Ekstr{\"o}m}}]{2016MNRAS.456.1803F}
{Frischknecht}, U., {Hirschi}, R., {Pignatari}, M., {et~al.} 2016, \mnras, 456,
  1803

\bibitem[{{Frischknecht} {et~al.}(2012){Frischknecht}, {Hirschi}, \&
  {Thielemann}}]{2012A&A...538L...2F}
{Frischknecht}, U., {Hirschi}, R., \& {Thielemann}, F.~K. 2012, \aap, 538, L2

\bibitem[{{Fu} {et~al.}(2018){Fu}, {Bressan}, {Marigo}, {Girardi},
  {Montalb{\'a}n}, {Chen}, \& {Nanni}}]{2018MNRAS.476..496F}
{Fu}, X., {Bressan}, A., {Marigo}, P., {et~al.} 2018, \mnras, 476, 496

\bibitem[{{Gallino} {et~al.}(1998){Gallino}, {Arlandini}, {Busso}, {Lugaro},
  {Travaglio}, {Straniero}, {Chieffi}, \& {Limongi}}]{1998ApJ...497..388G}
{Gallino}, R., {Arlandini}, C., {Busso}, M., {et~al.} 1998, \apj, 497, 388

\bibitem[{{Georgy} {et~al.}(2024){Georgy}, {Rizzuti}, {Hirschi}, {Varma},
  {Arnett}, {Meakin}, {Mocak}, {Murphy}, \& {Rauscher}}]{2024MNRAS.531.4293G}
{Georgy}, C., {Rizzuti}, F., {Hirschi}, R., {et~al.} 2024, \mnras, 531, 4293

\bibitem[{{Gonz{\'a}lez Hern{\'a}ndez} {et~al.}(2020){Gonz{\'a}lez
  Hern{\'a}ndez}, {Aguado}, {Allende Prieto}, {Burgasser}, \&
  {Rebolo}}]{2020ApJ...889L..13G}
{Gonz{\'a}lez Hern{\'a}ndez}, J.~I., {Aguado}, D.~S., {Allende Prieto}, C.,
  {Burgasser}, A.~J., \& {Rebolo}, R. 2020, \apjl, 889, L13

\bibitem[{{Goto} {et~al.}(2003){Goto}, {Usuda}, {Takato}, {Hayashi},
  {Sakamoto}, {Gaessler}, {Hayano}, {Iye}, {Kamata}, {Kanzawa}, {Kobayashi},
  {Minowa}, {Nedachi}, {Oya}, {Pyo}, {Saint-Jacques}, {Suto}, {Takami},
  {Terada}, \& {Mitchell}}]{2003ApJ...598.1038G}
{Goto}, M., {Usuda}, T., {Takato}, N., {et~al.} 2003, \apj, 598, 1038

\bibitem[{{Grisoni} {et~al.}(2020){Grisoni}, {Romano}, {Spitoni}, {Matteucci},
  {Ryde}, \& {J{\"o}nsson}}]{2020MNRAS.498.1252G}
{Grisoni}, V., {Romano}, D., {Spitoni}, E., {et~al.} 2020, \mnras, 498, 1252

\bibitem[{{Hansen} {et~al.}(2015){Hansen}, {Hansen}, {Christlieb}, {Beers},
  {Yong}, {Bessell}, {Frebel}, {Garc{\'\i}a P{\'e}rez}, {Placco}, {Norris}, \&
  {Asplund}}]{2015ApJ...807..173H}
{Hansen}, T., {Hansen}, C.~J., {Christlieb}, N., {et~al.} 2015, \apj, 807, 173

\bibitem[{{Heger} \& {Woosley}(2010)}]{2010ApJ...724..341H}
{Heger}, A. \& {Woosley}, S.~E. 2010, \apj, 724, 341

\bibitem[{{Herwig} {et~al.}(2011){Herwig}, {Pignatari}, {Woodward}, {Porter},
  {Rockefeller}, {Fryer}, {Bennett}, \& {Hirschi}}]{2011ApJ...727...89H}
{Herwig}, F., {Pignatari}, M., {Woodward}, P.~R., {et~al.} 2011, \apj, 727, 89

\bibitem[{{Herwig} {et~al.}(2014){Herwig}, Woodward, Lin, Knox, \&
  Fryer}]{Herwig_2014}
{Herwig}, F., Woodward, P.~R., Lin, P.-H., Knox, M., \& Fryer, C. 2014, \apjl,
  792, L3

\bibitem[{{Hirschi}(2007)}]{2007A&A...461..571H}
{Hirschi}, R. 2007, \aap, 461, 571

\bibitem[{{Ishimaru} \& {Wanajo}(1999)}]{1999ApJ...511L..33I}
{Ishimaru}, Y. \& {Wanajo}, S. 1999, \apjl, 511, L33

\bibitem[{{Israelian} {et~al.}(2004){Israelian}, {Ecuvillon}, {Rebolo},
  {Garc{\'\i}a-L{\'o}pez}, {Bonifacio}, \& {Molaro}}]{2004A&A...421..649I}
{Israelian}, G., {Ecuvillon}, A., {Rebolo}, R., {et~al.} 2004, \aap, 421, 649

\bibitem[{{Jones} {et~al.}(2016){Jones}, {Ritter}, {Herwig}, {Fryer},
  {Pignatari}, {Bertolli}, \& {Paxton}}]{2016MNRAS.455.3848J}
{Jones}, S., {Ritter}, C., {Herwig}, F., {et~al.} 2016, \mnras, 455, 3848

\bibitem[{{Karakas}(2010)}]{2010MNRAS.403.1413K}
{Karakas}, A.~I. 2010, \mnras, 403, 1413

\bibitem[{{Karlsson} \& {Gustafsson}(2005)}]{2005A&A...436..879K}
{Karlsson}, T. \& {Gustafsson}, B. 2005, \aap, 436, 879

\bibitem[{{Keller} {et~al.}(2014){Keller}, {Bessell}, {Frebel}, {Casey},
  {Asplund}, {Jacobson}, {Lind}, {Norris}, {Yong}, {Heger}, {Magic}, {da
  Costa}, {Schmidt}, \& {Tisserand}}]{2014Natur.506..463K}
{Keller}, S.~C., {Bessell}, M.~S., {Frebel}, A., {et~al.} 2014, \nat, 506, 463

\bibitem[{{Keller} {et~al.}(2007){Keller}, {Schmidt}, {Bessell}, {Conroy},
  {Francis}, {Granlund}, {Kowald}, {Oates}, {Martin-Jones}, {Preston},
  {Tisserand}, {Vaccarella}, \& {Waterson}}]{2007PASA...24....1K}
{Keller}, S.~C., {Schmidt}, B.~P., {Bessell}, M.~S., {et~al.} 2007, \pasa, 24,
  1

\bibitem[{{Kobayashi}(2022)}]{2022IAUS..366...63K}
{Kobayashi}, C. 2022, in IAU Symposium, Vol. 366, The Origin of Outflows in
  Evolved Stars, ed. L.~{Decin}, A.~{Zijlstra}, \& C.~{Gielen}, 63--82

\bibitem[{Kozyreva {et~al.}(2022)Kozyreva, Janka, Kresse, Taubenberger, \&
  Baklanov}]{10.1093/mnras/stac1518}
Kozyreva, A., Janka, H.-T., Kresse, D., Taubenberger, S., \& Baklanov, P. 2022,
  Monthly Notices of the Royal Astronomical Society, 514, 4173

\bibitem[{{Kroupa}(2001)}]{2001MNRAS.322..231K}
{Kroupa}, P. 2001, \mnras, 322, 231

\bibitem[{{Lai} {et~al.}(2008){Lai}, {Bolte}, {Johnson}, {Lucatello}, {Heger},
  \& {Woosley}}]{2008ApJ...681.1524L}
{Lai}, D.~K., {Bolte}, M., {Johnson}, J.~A., {et~al.} 2008, \apj, 681, 1524

\bibitem[{{Limongi} \& {Chieffi}(2012)}]{2012ApJS..199...38L}
{Limongi}, M. \& {Chieffi}, A. 2012, \apjs, 199, 38

\bibitem[{Limongi \& Chieffi(2018)}]{Limongi_2018}
Limongi, M. \& Chieffi, A. 2018, The Astrophysical Journal Supplement Series,
  237, 13

\bibitem[{{Lind} \& {Amarsi}(2024)}]{2024arXiv240100697L}
{Lind}, K. \& {Amarsi}, A.~M. 2024, arXiv e-prints, arXiv:2401.00697

\bibitem[{{Lucatello} {et~al.}(2005){Lucatello}, {Tsangarides}, {Beers},
  {Carretta}, {Gratton}, \& {Ryan}}]{2005ApJ...625..825L}
{Lucatello}, S., {Tsangarides}, S., {Beers}, T.~C., {et~al.} 2005, \apj, 625,
  825

\bibitem[{{Maeder} \& {Meynet}(1989)}]{1989A&A...210..155M}
{Maeder}, A. \& {Meynet}, G. 1989, \aap, 210, 155

\bibitem[{{Maeder} \& {Meynet}(2015)}]{2015A&A...580A..32M}
{Maeder}, A. \& {Meynet}, G. 2015, \aap, 580, A32

\bibitem[{{Marini} {et~al.}(2023){Marini}, {Ventura}, {Tailo}, {Ventura},
  {Dell'Agli}, \& {Castellani}}]{2023A&A...676A..19M}
{Marini}, E., {Ventura}, C., {Tailo}, M., {et~al.} 2023, \aap, 676, A19

\bibitem[{{Masseron} {et~al.}(2010){Masseron}, {Johnson}, {Plez}, {van Eck},
  {Primas}, {Goriely}, \& {Jorissen}}]{2010A&A...509A..93M}
{Masseron}, T., {Johnson}, J.~A., {Plez}, B., {et~al.} 2010, \aap, 509, A93

\bibitem[{{Matsuno} {et~al.}(2017){Matsuno}, {Aoki}, {Suda}, \&
  {Li}}]{2017PASJ...69...24M}
{Matsuno}, T., {Aoki}, W., {Suda}, T., \& {Li}, H. 2017, \pasj, 69, 24

\bibitem[{{Matteucci}(1986)}]{1986MNRAS.221..911M}
{Matteucci}, F. 1986, \mnras, 221, 911

\bibitem[{{Matteucci} {et~al.}(2014){Matteucci}, {Romano}, {Arcones},
  {Korobkin}, \& {Rosswog}}]{2014MNRAS.438.2177M}
{Matteucci}, F., {Romano}, D., {Arcones}, A., {Korobkin}, O., \& {Rosswog}, S.
  2014, \mnras, 438, 2177

\bibitem[{{Meynet} {et~al.}(2010){Meynet}, {Hirschi}, {Ekstrom}, {Maeder},
  {Georgy}, {Eggenberger}, \& {Chiappini}}]{2010A&A...521A..30M}
{Meynet}, G., {Hirschi}, R., {Ekstrom}, S., {et~al.} 2010, \aap, 521, A30

\bibitem[{{Meynet} \& {Maeder}(2002)}]{2002A&A...390..561M}
{Meynet}, G. \& {Maeder}, A. 2002, \aap, 390, 561

\bibitem[{{Moc{\'a}k} {et~al.}(2018){Moc{\'a}k}, {Meakin}, {Campbell}, \&
  {Arnett}}]{2018MNRAS.481.2918M}
{Moc{\'a}k}, M., {Meakin}, C., {Campbell}, S.~W., \& {Arnett}, W.~D. 2018,
  \mnras, 481, 2918

\bibitem[{{Moc{\'a}k} {et~al.}(2011){Moc{\'a}k}, {Siess}, \&
  {M{\"u}ller}}]{2011A&A...533A..53M}
{Moc{\'a}k}, M., {Siess}, L., \& {M{\"u}ller}, E. 2011, \aap, 533, A53

\bibitem[{{Mohorian} {et~al.}(2024){Mohorian}, {Kamath}, {Menon}, {Ventura},
  {Van Winckel}, {Garc{\'\i}a-Hern{\'a}ndez}, \&
  {Masseron}}]{2024MNRAS.530..761M}
{Mohorian}, M., {Kamath}, D., {Menon}, M., {et~al.} 2024, \mnras, 530, 761

\bibitem[{{Molaro} {et~al.}(2023){Molaro}, {Aguado, D. S.}, {Caffau, E.},
  {Allende Prieto, C.}, {Bonifacio, P.}, {González Hernández, J. I.},
  {Rebolo, R.}, {Zapatero Osorio, M. R.}, {Cristiani, S.}, {Pepe, F.}, {Santos,
  N. C.}, {Alibert, Y.}, {Cupani, G.}, {Di Marcantonio, P.}, {D’Odorico, V.},
  {Lovis, C.}, {Martins, C. J. A. P.}, {Milaković, D.}, {Murphy, M. T.},
  {Nunes, N. J.}, {Schmidt, T. M.}, {Sousa, S.}, {Sozzetti, A.}, \& {Suárez
  Mascareño, A.}}]{Molaro}
{Molaro}, P., {Aguado, D. S.}, {Caffau, E.}, {et~al.} 2023, \aap, 679, A72

\bibitem[{{Molaro} \& {Bonifacio}(1990)}]{1990A&A...236L...5M}
{Molaro}, P. \& {Bonifacio}, P. 1990, \aap, 236, L5

\bibitem[{{Molaro} \& {Castelli}(1990)}]{1990A&A...228..426M}
{Molaro}, P. \& {Castelli}, F. 1990, \aap, 228, 426

\bibitem[{{Molero} {et~al.}(2023){Molero}, {Magrini}, {Matteucci}, {Romano},
  {Palla}, {Cescutti}, {Viscasillas V{\'a}zquez}, \&
  {Spitoni}}]{2023MNRAS.523.2974M}
{Molero}, M., {Magrini}, L., {Matteucci}, F., {et~al.} 2023, \mnras, 523, 2974

\bibitem[{{Molero} {et~al.}(2021){Molero}, {Simonetti}, {Matteucci}, \& {della
  Valle}}]{2021MNRAS.500.1071M}
{Molero}, M., {Simonetti}, P., {Matteucci}, F., \& {della Valle}, M. 2021,
  \mnras, 500, 1071

\bibitem[{{M{\"u}ller} {et~al.}(2016){M{\"u}ller}, {Heger}, {Liptai}, \&
  {Cameron}}]{2016MNRAS.460..742M}
{M{\"u}ller}, B., {Heger}, A., {Liptai}, D., \& {Cameron}, J.~B. 2016, \mnras,
  460, 742

\bibitem[{{M{\"u}ller} \& {Janka}(2015)}]{2015MNRAS.448.2141M}
{M{\"u}ller}, B. \& {Janka}, H.~T. 2015, \mnras, 448, 2141

\bibitem[{{Nguyen} {et~al.}(2024){Nguyen}, {Bressan}, {Korn}, {Cescutti},
  {Costa}, {Addari}, {Girardi}, {Fu}, {Chen}, \&
  {Marigo}}]{2024arXiv240805039N}
{Nguyen}, C.~T., {Bressan}, A., {Korn}, A.~J., {et~al.} 2024, arXiv e-prints,
  arXiv:2408.05039

\bibitem[{{Nomoto} {et~al.}(2013){Nomoto}, {Kobayashi}, \&
  {Tominaga}}]{2013ARA&A..51..457N}
{Nomoto}, K., {Kobayashi}, C., \& {Tominaga}, N. 2013, \araa, 51, 457

\bibitem[{{Nordlander} {et~al.}(2019){Nordlander}, {Bessell}, {Da Costa},
  {Mackey}, {Asplund}, {Casey}, {Chiti}, {Ezzeddine}, {Frebel}, {Lind},
  {Marino}, {Murphy}, {Norris}, {Schmidt}, \& {Yong}}]{2019MNRAS.488L.109N}
{Nordlander}, T., {Bessell}, M.~S., {Da Costa}, G.~S., {et~al.} 2019, \mnras,
  488, L109

\bibitem[{{Norris} {et~al.}(2013){Norris}, {Bessell}, {Yong}, {Christlieb},
  {Barklem}, {Asplund}, {Murphy}, {Beers}, {Frebel}, \&
  {Ryan}}]{2013ApJ...762...25N}
{Norris}, J.~E., {Bessell}, M.~S., {Yong}, D., {et~al.} 2013, \apj, 762, 25

\bibitem[{{Norris} {et~al.}(1997){Norris}, {Ryan}, \&
  {Beers}}]{1997ApJ...489L.169N}
{Norris}, J.~E., {Ryan}, S.~G., \& {Beers}, T.~C. 1997, \apjl, 489, L169

\bibitem[{{Perego} {et~al.}(2015){Perego}, {Hempel}, {Fr{\"o}hlich}, {Ebinger},
  {Eichler}, {Casanova}, {Liebend{\"o}rfer}, \&
  {Thielemann}}]{2015ApJ...806..275P}
{Perego}, A., {Hempel}, M., {Fr{\"o}hlich}, C., {et~al.} 2015, \apj, 806, 275

\bibitem[{{Placco} {et~al.}(2014){Placco}, {Frebel}, {Beers}, \&
  {Stancliffe}}]{2014ApJ...797...21P}
{Placco}, V.~M., {Frebel}, A., {Beers}, T.~C., \& {Stancliffe}, R.~J. 2014,
  \apj, 797, 21

\bibitem[{{Prantzos} {et~al.}(2023){Prantzos}, {Abia}, {Chen}, {de Laverny},
  {Recio-Blanco}, {Athanassoula}, {Roberti}, {Vescovi}, {Limongi}, {Chieffi},
  \& {Cristallo}}]{2023MNRAS.523.2126P}
{Prantzos}, N., {Abia}, C., {Chen}, T., {et~al.} 2023, \mnras, 523, 2126

\bibitem[{{Prantzos} {et~al.}(2018){Prantzos}, {Abia}, {Limongi}, {Chieffi}, \&
  {Cristallo}}]{2018MNRAS.476.3432P}
{Prantzos}, N., {Abia}, C., {Limongi}, M., {Chieffi}, A., \& {Cristallo}, S.
  2018, \mnras, 476, 3432

\bibitem[{{Renzini} \& {Voli}(1981)}]{1981A&A....94..175R}
{Renzini}, A. \& {Voli}, M. 1981, \aap, 94, 175

\bibitem[{{Ritter} {et~al.}(2018){Ritter}, {Herwig}, {Jones}, {Pignatari},
  {Fryer}, \& {Hirschi}}]{2018MNRAS.480..538R}
{Ritter}, C., {Herwig}, F., {Jones}, S., {et~al.} 2018, \mnras, 480, 538

\bibitem[{{Rizzuti} {et~al.}(2019){Rizzuti}, {Cescutti}, {Matteucci},
  {Chieffi}, {Hirschi}, \& {Limongi}}]{Rizzuti2019}
{Rizzuti}, F., {Cescutti}, G., {Matteucci}, F., {et~al.} 2019, \mnras, 489,
  5244

\bibitem[{{Rizzuti} {et~al.}(2021){Rizzuti}, {Cescutti}, {Matteucci},
  {Chieffi}, {Hirschi}, {Limongi}, \& {Saro}}]{Rizzuti2021}
{Rizzuti}, F., {Cescutti}, G., {Matteucci}, F., {et~al.} 2021, \mnras, 502,
  2495

\bibitem[{Rizzuti {et~al.}(2023)Rizzuti, Hirschi, Arnett, Georgy, Meakin,
  Murphy, Rauscher, \& Varma}]{2023Rizzuti}
Rizzuti, F., Hirschi, R., Arnett, W.~D., {et~al.} 2023, \mnras, 523, 2317

\bibitem[{Rizzuti {et~al.}(2022)Rizzuti, Hirschi, Georgy, Arnett, Meakin, \&
  Murphy}]{2022Rizzuti}
Rizzuti, F., Hirschi, R., Georgy, C., {et~al.} 2022, \mnras, 515, 4013

\bibitem[{{Rizzuti} {et~al.}(2024){Rizzuti}, {Hirschi}, {Varma}, {Arnett},
  {Georgy}, {Meakin}, {Moc{\'a}k}, {Murphy}, \&
  {Rauscher}}]{2024MNRAS.533..687R}
{Rizzuti}, F., {Hirschi}, R., {Varma}, V., {et~al.} 2024, \mnras, 533, 687

\bibitem[{Roberti {et~al.}(2024)Roberti, Limongi, \& Chieffi}]{Roberti_2024}
Roberti, L., Limongi, M., \& Chieffi, A. 2024, The Astrophysical Journal
  Supplement Series, 270, 28

\bibitem[{{Rockosi} {et~al.}(2022){Rockosi}, {Lee}, {Morrison}, {Yanny},
  {Johnson}, {Lucatello}, {Sobeck}, {Beers}, {Allende Prieto}, {An}, {Bizyaev},
  {Blanton}, {Casagrande}, {Eisenstein}, {Gould}, {Gunn}, {Harding}, {Ivans},
  {Jacobson}, {Janesh}, {Knapp}, {Kollmeier}, {L{\'e}pine},
  {L{\'o}pez-Corredoira}, {Ma}, {Newberg}, {Pan}, {Prchlik}, {Sayers},
  {Schlesinger}, {Simmerer}, \& {Weinberg}}]{2022ApJS..259...60R}
{Rockosi}, C.~M., {Lee}, Y.~S., {Morrison}, H.~L., {et~al.} 2022, \apjs, 259,
  60

\bibitem[{{Romano}(2022)}]{2022A&ARv..30....7R}
{Romano}, D. 2022, \aapr, 30, 7

\bibitem[{{Romano} \& {Matteucci}(2003)}]{2003MNRAS.342..185R}
{Romano}, D. \& {Matteucci}, F. 2003, \mnras, 342, 185

\bibitem[{{Romano} {et~al.}(2019){Romano}, {Matteucci}, {Zhang}, {Ivison}, \&
  {Ventura}}]{2019MNRAS.490.2838R}
{Romano}, D., {Matteucci}, F., {Zhang}, Z.-Y., {Ivison}, R.~J., \& {Ventura},
  P. 2019, \mnras, 490, 2838

\bibitem[{{Romano} {et~al.}(2017){Romano}, {Matteucci}, {Zhang},
  {Papadopoulos}, \& {Ivison}}]{2017MNRAS.470..401R}
{Romano}, D., {Matteucci}, F., {Zhang}, Z.~Y., {Papadopoulos}, P.~P., \&
  {Ivison}, R.~J. 2017, \mnras, 470, 401

\bibitem[{{Rossi} {et~al.}(2024){Rossi}, {Romano}, {Mucciarelli}, {Ceccarelli},
  {Massari}, \& {Zamorani}}]{2024A&A...691A.284R}
{Rossi}, M., {Romano}, D., {Mucciarelli}, A., {et~al.} 2024, \aap, 691, A284

\bibitem[{{Rossi} {et~al.}(2021){Rossi}, {Salvadori}, \&
  {Sk{\'u}lad{\'o}ttir}}]{2021MNRAS.503.6026R}
{Rossi}, M., {Salvadori}, S., \& {Sk{\'u}lad{\'o}ttir}, {\'A}. 2021, \mnras,
  503, 6026

\bibitem[{{Scalo}(1986)}]{1986FCPh...11....1S}
{Scalo}, J.~M. 1986, \fcp, 11, 1

\bibitem[{{Schneider} {et~al.}(2018){Schneider}, {Sana}, {Evans},
  {Bestenlehner}, {Castro}, {Fossati}, {Gr{\"a}fener}, {Langer},
  {Ram{\'\i}rez-Agudelo}, {Sab{\'\i}n-Sanjuli{\'a}n}, {Sim{\'o}n-D{\'\i}az},
  {Tramper}, {Crowther}, {de Koter}, {de Mink}, {Dufton}, {Garcia}, {Gieles},
  {H{\'e}nault-Brunet}, {Herrero}, {Izzard}, {Kalari}, {Lennon}, {Ma{\'\i}z
  Apell{\'a}niz}, {Markova}, {Najarro}, {Podsiadlowski}, {Puls}, {Taylor}, {van
  Loon}, {Vink}, \& {Norman}}]{2018Sci...359...69S}
{Schneider}, F.~R.~N., {Sana}, H., {Evans}, C.~J., {et~al.} 2018, Science, 359,
  69

\bibitem[{Scott {et~al.}(2021)Scott, Hirschi, Georgy, Arnett, Meakin, Kaiser,
  Ekström, \& Yusof}]{2021Scott}
Scott, L. J.~A., Hirschi, R., Georgy, C., {et~al.} 2021, \mnras, 503, 4208

\bibitem[{{Simonetti} {et~al.}(2019){Simonetti}, {Matteucci}, {Greggio}, \&
  {Cescutti}}]{2019MNRAS.486.2896S}
{Simonetti}, P., {Matteucci}, F., {Greggio}, L., \& {Cescutti}, G. 2019,
  \mnras, 486, 2896

\bibitem[{{Sivarani} {et~al.}(2006){Sivarani}, {Beers}, {Bonifacio}, {Molaro},
  {Cayrel}, {Herwig}, {Spite}, {Spite}, {Plez}, {Andersen}, {Barbuy},
  {Depagne}, {Hill}, {Fran{\c{c}}ois}, {Nordstr{\"o}m}, \&
  {Primas}}]{2006A&A...459..125S}
{Sivarani}, T., {Beers}, T.~C., {Bonifacio}, P., {et~al.} 2006, \aap, 459, 125

\bibitem[{Smartt {et~al.}(2009)Smartt, Eldridge, Crockett, \& Maund}]{Smartt}
Smartt, S.~J., Eldridge, J.~J., Crockett, R.~M., \& Maund, J.~R. 2009, Monthly
  Notices of the Royal Astronomical Society, 395, 1409

\bibitem[{Sollerman {et~al.}(1998)Sollerman, Cumming, \&
  Lundqvist}]{Sollerman_1998}
Sollerman, J., Cumming, R.~J., \& Lundqvist, P. 1998, The Astrophysical
  Journal, 493, 933

\bibitem[{{Spite} {et~al.}(2013){Spite}, {Caffau}, {Bonifacio}, {Spite},
  {Ludwig}, {Plez}, \& {Christlieb}}]{2013A&A...552A.107S}
{Spite}, M., {Caffau}, E., {Bonifacio}, P., {et~al.} 2013, \aap, 552, A107

\bibitem[{{Spite} {et~al.}(2005){Spite}, {Cayrel}, {Plez}, {Hill}, {Spite},
  {Depagne}, {Fran{\c{c}}ois}, {Bonifacio}, {Barbuy}, {Beers}, {Andersen},
  {Molaro}, {Nordstr{\"o}m}, \& {Primas}}]{2005A&A...430..655S}
{Spite}, M., {Cayrel}, R., {Plez}, B., {et~al.} 2005, \aap, 430, 655

\bibitem[{{Spite} {et~al.}(2006){Spite}, {Cayrel, R.}, {Hill, V.}, {Spite, F.},
  {François, P.}, {Plez, B.}, {Bonifacio, P.}, {Molaro, P.}, {Depagne, E.},
  {Andersen, J.}, {Barbuy, B.}, {Beers, T. C.}, {Nordström, B.}, \& {Primas,
  F.}}]{Spite}
{Spite}, M., {Cayrel, R.}, {Hill, V.}, {et~al.} 2006, \aap, 455, 291

\bibitem[{{Spite} {et~al.}(2021){Spite}, {Spite}, \&
  {Barbuy}}]{2021A&A...652A..97S}
{Spite}, M., {Spite}, F., \& {Barbuy}, B. 2021, \aap, 652, A97

\bibitem[{{Stancliffe} {et~al.}(2011){Stancliffe}, {Dearborn}, {Lattanzio},
  {Heap}, \& {Campbell}}]{2011ApJ...742..121S}
{Stancliffe}, R.~J., {Dearborn}, D. S.~P., {Lattanzio}, J.~C., {Heap}, S.~A.,
  \& {Campbell}, S.~W. 2011, \apj, 742, 121

\bibitem[{{Starkenburg} {et~al.}(2017){Starkenburg}, {Martin}, {Youakim},
  {Aguado}, {Allende Prieto}, {Arentsen}, {Bernard}, {Bonifacio}, {Caffau},
  {Carlberg}, {C{\^o}t{\'e}}, {Fouesneau}, {Fran{\c{c}}ois}, {Franke},
  {Gonz{\'a}lez Hern{\'a}ndez}, {Gwyn}, {Hill}, {Ibata}, {Jablonka},
  {Longeard}, {McConnachie}, {Navarro}, {S{\'a}nchez-Janssen}, {Tolstoy}, \&
  {Venn}}]{2017MNRAS.471.2587S}
{Starkenburg}, E., {Martin}, N., {Youakim}, K., {et~al.} 2017, \mnras, 471,
  2587

\bibitem[{{Starkenburg} {et~al.}(2014){Starkenburg}, {Shetrone}, {McConnachie},
  \& {Venn}}]{2014MNRAS.441.1217S}
{Starkenburg}, E., {Shetrone}, M.~D., {McConnachie}, A.~W., \& {Venn}, K.~A.
  2014, \mnras, 441, 1217

\bibitem[{Stockinger {et~al.}(2020)Stockinger, Janka, Kresse, Melson, Ertl,
  Gabler, Gessner, Wongwathanarat, Tolstov, Leung, Nomoto, \&
  Heger}]{10.1093/mnras/staa1691}
Stockinger, G., Janka, H.-T., Kresse, D., {et~al.} 2020, Monthly Notices of the
  Royal Astronomical Society, 496, 2039

\bibitem[{{Sukhbold} {et~al.}(2016){Sukhbold}, {Ertl}, {Woosley}, {Brown}, \&
  {Janka}}]{2016ApJ...821...38S}
{Sukhbold}, T., {Ertl}, T., {Woosley}, S.~E., {Brown}, J.~M., \& {Janka}, H.~T.
  2016, \apj, 821, 38

\bibitem[{{Sukhbold} \& {Woosley}(2014)}]{2014ApJ...783...10S}
{Sukhbold}, T. \& {Woosley}, S.~E. 2014, \apj, 783, 10

\bibitem[{{Thornton} {et~al.}(1998){Thornton}, {Gaudlitz}, {Janka}, \&
  {Steinmetz}}]{1998ApJ...500...95T}
{Thornton}, K., {Gaudlitz}, M., {Janka}, H.~T., \& {Steinmetz}, M. 1998, \apj,
  500, 95

\bibitem[{{Tominaga} {et~al.}(2014){Tominaga}, {Iwamoto}, \&
  {Nomoto}}]{2014ApJ...785...98T}
{Tominaga}, N., {Iwamoto}, N., \& {Nomoto}, K. 2014, \apj, 785, 98

\bibitem[{Tominaga {et~al.}(2014)Tominaga, Iwamoto, \& Nomoto}]{Tominaga_2014}
Tominaga, N., Iwamoto, N., \& Nomoto, K. 2014, The Astrophysical Journal, 785,
  98

\bibitem[{{Tominaga} {et~al.}(2007){Tominaga}, {Umeda}, \&
  {Nomoto}}]{2007ApJ...660..516T}
{Tominaga}, N., {Umeda}, H., \& {Nomoto}, K. 2007, \apj, 660, 516

\bibitem[{{Travaglio} {et~al.}(2001){Travaglio}, {Galli}, \&
  {Burkert}}]{2001ApJ...547..217T}
{Travaglio}, C., {Galli}, D., \& {Burkert}, A. 2001, \apj, 547, 217

\bibitem[{{Tsiatsiou} {et~al.}(2024){Tsiatsiou}, {Sibony}, {Nandal},
  {Sciarini}, {Hirai}, {Ekstr{\"o}m}, {Farrell}, {Murphy}, {Choplin},
  {Hirschi}, {Chiappini}, {Liu}, {Bromm}, {Groh}, \&
  {Meynet}}]{2024A&A...687A.307T}
{Tsiatsiou}, S., {Sibony}, Y., {Nandal}, D., {et~al.} 2024, \aap, 687, A307

\bibitem[{{Tsujimoto} {et~al.}(1999){Tsujimoto}, {Shigeyama}, \&
  {Yoshii}}]{1999ApJ...519L..63T}
{Tsujimoto}, T., {Shigeyama}, T., \& {Yoshii}, Y. 1999, \apjl, 519, L63

\bibitem[{Turatto {et~al.}(1998)Turatto, Mazzali, Young, Nomoto, Iwamoto,
  Benetti, Cappellaro, Danziger, de~Mello, Phillips, Suntzeff, Clocchiatti,
  Piemonte, Leibundgut, Covarrubias, Maza, \& Sollerman}]{Turatto_1998}
Turatto, M., Mazzali, P.~A., Young, T.~R., {et~al.} 1998, The Astrophysical
  Journal, 498, L129

\bibitem[{{Umeda} \& {Nomoto}(2003)}]{2003Natur.422..871U}
{Umeda}, H. \& {Nomoto}, K. 2003, \nat, 422, 871

\bibitem[{{Umeda} \& {Nomoto}(2005)}]{2005ApJ...619..427U}
{Umeda}, H. \& {Nomoto}, K. 2005, \apj, 619, 427

\bibitem[{{Varma} {et~al.}(2023){Varma}, {M{\"u}ller}, \&
  {Schneider}}]{2023MNRAS.518.3622V}
{Varma}, V., {M{\"u}ller}, B., \& {Schneider}, F. R.~N. 2023, \mnras, 518, 3622

\bibitem[{{Wehmeyer} {et~al.}(2015){Wehmeyer}, {Pignatari}, \&
  {Thielemann}}]{2015MNRAS.452.1970W}
{Wehmeyer}, B., {Pignatari}, M., \& {Thielemann}, F.~K. 2015, \mnras, 452, 1970

\bibitem[{{Woosley} \& {Weaver}(1982)}]{1982ASIC...90...79W}
{Woosley}, S.~E. \& {Weaver}, T.~A. 1982, in NATO Advanced Study Institute
  (ASI) Series C, Vol.~90, Supernovae: A Survey of Current Research, ed. M.~J.
  {Rees} \& R.~J. {Stoneham}, 79

\bibitem[{{Woosley} \& {Weaver}(1995)}]{1995ApJS..101..181W}
{Woosley}, S.~E. \& {Weaver}, T.~A. 1995, \apjs, 101, 181

\bibitem[{Yadav {et~al.}(2020)Yadav, Müller, Janka, Melson, \&
  Heger}]{Yadav_2020}
Yadav, N., Müller, B., Janka, H.~T., Melson, T., \& Heger, A. 2020, \apj, 890,
  94

\bibitem[{{Yanny} {et~al.}(2009){Yanny}, {Rockosi}, {Newberg}, {Knapp},
  {Adelman-McCarthy}, {Alcorn}, {Allam}, {Allende Prieto}, {An}, {Anderson},
  {Anderson}, {Bailer-Jones}, {Bastian}, {Beers}, {Bell}, {Belokurov},
  {Bizyaev}, {Blythe}, {Bochanski}, {Boroski}, {Brinchmann}, {Brinkmann},
  {Brewington}, {Carey}, {Cudworth}, {Evans}, {Evans}, {Gates}, {G{\"a}nsicke},
  {Gillespie}, {Gilmore}, {Nebot Gomez-Moran}, {Grebel}, {Greenwell}, {Gunn},
  {Jordan}, {Jordan}, {Harding}, {Harris}, {Hendry}, {Holder}, {Ivans},
  {Ivezi{\v{c}}}, {Jester}, {Johnson}, {Kent}, {Kleinman}, {Kniazev},
  {Krzesinski}, {Kron}, {Kuropatkin}, {Lebedeva}, {Lee}, {French Leger},
  {L{\'e}pine}, {Levine}, {Lin}, {Long}, {Loomis}, {Lupton}, {Malanushenko},
  {Malanushenko}, {Margon}, {Martinez-Delgado}, {McGehee}, {Monet}, {Morrison},
  {Munn}, {Neilsen}, {Nitta}, {Norris}, {Oravetz}, {Owen}, {Padmanabhan},
  {Pan}, {Peterson}, {Pier}, {Platson}, {Re Fiorentin}, {Richards}, {Rix},
  {Schlegel}, {Schneider}, {Schreiber}, {Schwope}, {Sibley}, {Simmons},
  {Snedden}, {Allyn Smith}, {Stark}, {Stauffer}, {Steinmetz}, {Stoughton},
  {SubbaRao}, {Szalay}, {Szkody}, {Thakar}, {Sivarani}, {Tucker}, {Uomoto},
  {Vanden Berk}, {Vidrih}, {Wadadekar}, {Watters}, {Wilhelm}, {Wyse}, {Yarger},
  \& {Zucker}}]{2009AJ....137.4377Y}
{Yanny}, B., {Rockosi}, C., {Newberg}, H.~J., {et~al.} 2009, \aj, 137, 4377

\bibitem[{{Yoon} {et~al.}(2018){Yoon}, {Beers}, {Dietz}, {Lee}, {Placco}, {Da
  Costa}, {Keller}, {Owen}, \& {Sharma}}]{2018ApJ...861..146Y}
{Yoon}, J., {Beers}, T.~C., {Dietz}, S., {et~al.} 2018, \apj, 861, 146

\bibitem[{{York} {et~al.}(2000){York}, {Adelman}, {Anderson}, {Anderson},
  {Annis}, {Bahcall}, {Bakken}, {Barkhouser}, {Bastian}, {Berman}, {Boroski},
  {Bracker}, {Briegel}, {Briggs}, {Brinkmann}, {Brunner}, {Burles}, {Carey},
  {Carr}, {Castander}, {Chen}, {Colestock}, {Connolly}, {Crocker}, {Csabai},
  {Czarapata}, {Davis}, {Doi}, {Dombeck}, {Eisenstein}, {Ellman}, {Elms},
  {Evans}, {Fan}, {Federwitz}, {Fiscelli}, {Friedman}, {Frieman}, {Fukugita},
  {Gillespie}, {Gunn}, {Gurbani}, {de Haas}, {Haldeman}, {Harris}, {Hayes},
  {Heckman}, {Hennessy}, {Hindsley}, {Holm}, {Holmgren}, {Huang}, {Hull},
  {Husby}, {Ichikawa}, {Ichikawa}, {Ivezi{\'c}}, {Kent}, {Kim}, {Kinney},
  {Klaene}, {Kleinman}, {Kleinman}, {Knapp}, {Korienek}, {Kron}, {Kunszt},
  {Lamb}, {Lee}, {Leger}, {Limmongkol}, {Lindenmeyer}, {Long}, {Loomis},
  {Loveday}, {Lucinio}, {Lupton}, {MacKinnon}, {Mannery}, {Mantsch}, {Margon},
  {McGehee}, {McKay}, {Meiksin}, {Merelli}, {Monet}, {Munn}, {Narayanan},
  {Nash}, {Neilsen}, {Neswold}, {Newberg}, {Nichol}, {Nicinski}, {Nonino},
  {Okada}, {Okamura}, {Ostriker}, {Owen}, {Pauls}, {Peoples}, {Peterson},
  {Petravick}, {Pier}, {Pope}, {Pordes}, {Prosapio}, {Rechenmacher}, {Quinn},
  {Richards}, {Richmond}, {Rivetta}, {Rockosi}, {Ruthmansdorfer}, {Sandford},
  {Schlegel}, {Schneider}, {Sekiguchi}, {Sergey}, {Shimasaku}, {Siegmund},
  {Smee}, {Smith}, {Snedden}, {Stone}, {Stoughton}, {Strauss}, {Stubbs},
  {SubbaRao}, {Szalay}, {Szapudi}, {Szokoly}, {Thakar}, {Tremonti}, {Tucker},
  {Uomoto}, {Vanden Berk}, {Vogeley}, {Waddell}, {Wang}, {Watanabe},
  {Weinberg}, {Yanny}, {Yasuda}, \& {SDSS Collaboration}}]{2000AJ....120.1579Y}
{York}, D.~G., {Adelman}, J., {Anderson}, Jr., J.~E., {et~al.} 2000, \aj, 120,
  1579

\bibitem[{{Zhang} {et~al.}(2018){Zhang}, {Romano}, {Ivison}, {Papadopoulos}, \&
  {Matteucci}}]{2018Natur.558..260Z}
{Zhang}, Z.-Y., {Romano}, D., {Ivison}, R.~J., {Papadopoulos}, P.~P., \&
  {Matteucci}, F. 2018, \nat, 558, 260

\bibitem[{{Zhao} {et~al.}(2016){Zhao}, {Mashonkina}, {Yan}, {Alexeeva},
  {Kobayashi}, {Pakhomov}, {Shi}, {Sitnova}, {Tan}, {Zhang}, {Zhang}, {Zhou},
  {Bolte}, {Chen}, {Li}, {Liu}, \& {Zhai}}]{2016ApJ...833..225Z}
{Zhao}, G., {Mashonkina}, L., {Yan}, H.~L., {et~al.} 2016, \apj, 833, 225

\end{thebibliography}
%
\begin{appendix} 
   
\section{Kiel diagram of the observations}
\begin{figure*}
   \centering
   \includegraphics[width=\hsize]{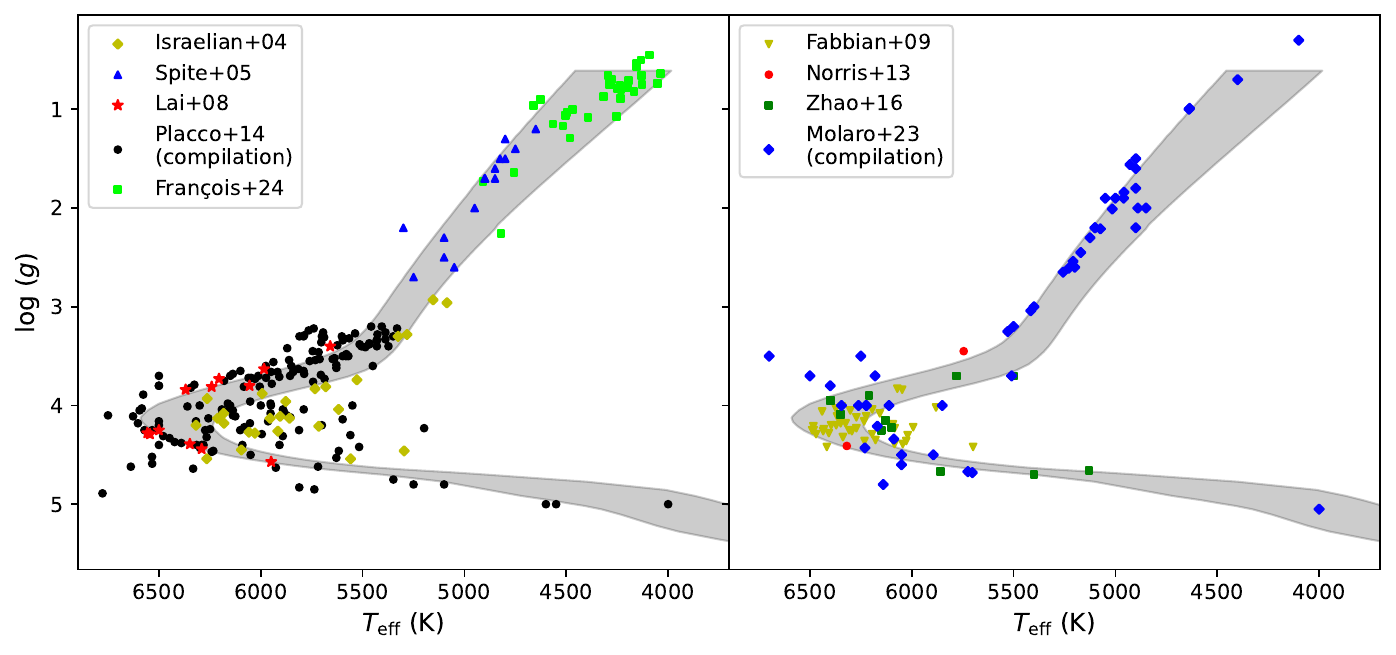}
      \caption{Kiel diagram ($\log(g)$ versus $T_\text{eff}$) for the observations presented in Section \ref{sec:obs}. The shaded areas are 12 Gyr isochrones obtained from \texttt{PARSEC} models, with metallicity between \feh{=-2.4} ($Z=0.0001$) and \feh{=-1.2} ($Z=0.0015$).}\label{fig:iso}
   \end{figure*}

Evolutionary effects during the lifetime of observed stars could modify the original surface abundances of light elements. In order to compare the chemical evolution models to original surface abundances, we made a selection on the observation (see Section \ref{sec:obs}). We present in Fig.~\ref{fig:iso} the $\log(g)$ versus $T_\text{eff}$ of all the stars considered in this work. We compare them to isochrones obtained from \texttt{PARSEC} models \citep{2012MNRAS.427..127B, 2014MNRAS.444.2525C, 2018MNRAS.476..496F} for ages of 12 Gyr and metallicity between $-2.4<$ \feh{<-1.2}; unfortunately, no lower metallicity is available. Most stars have been selected to be dwarfs with $\log(g)\geq3.2$, to avoid internal mixing effects. The only stars with $\log(g)<3.2$ are from \cite{2005A&A...430..655S} who used Li to select unmixed giants, \cite{2024A&A...686A.295F} from which we only used Sr and Ba abundances, and \cite{Molaro} where we make a distinction between dwarfs and giants when comparing to the models.

\section{Age-metallicity relation}
\begin{figure}
   \centering
   \includegraphics[width=\columnwidth]{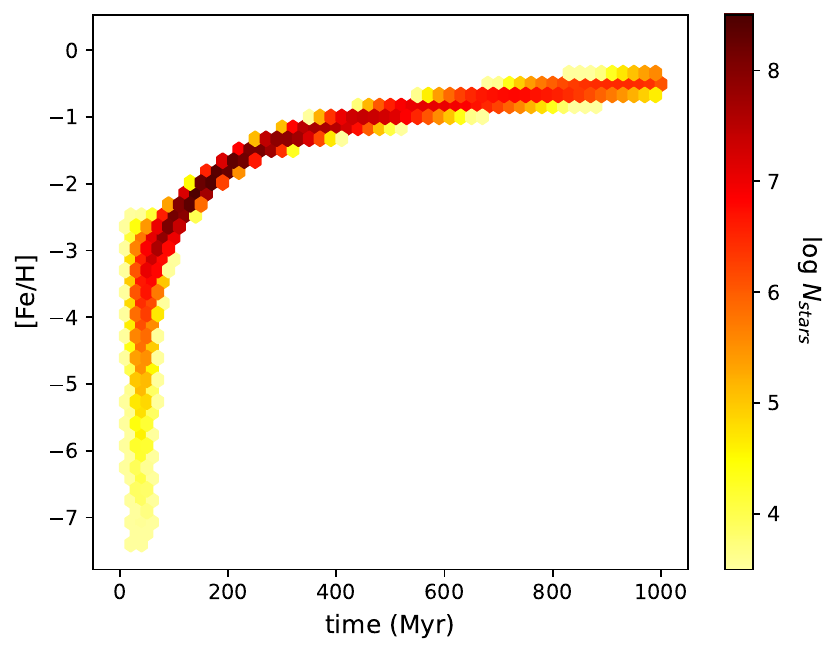}
      \caption{Age-metallicity relation ([Fe/H] versus time) in the stochastic model of Fig.~\ref{fig:SFR} as colour map.}\label{fig:age}
   \end{figure}

In order to show the impact of stars of different mass on the nucleosynthesis, we show in Fig.~\ref{fig:age} the relation between [Fe/H] and time in our stochastic simulation. The dispersion is very narrow and the trend is monotonically increasing. In particular, we notice that the first massive AGB stars $\lesssim6 \text{ M}_\odot$ that die after $\gtrsim60$ Myr are only in large enough number to be impactful after \feh{>-3}, and cannot die before \feh{=-5}.

\section{Changing the initial mass function}
\begin{figure}
   \centering
   \includegraphics[width=\hsize]{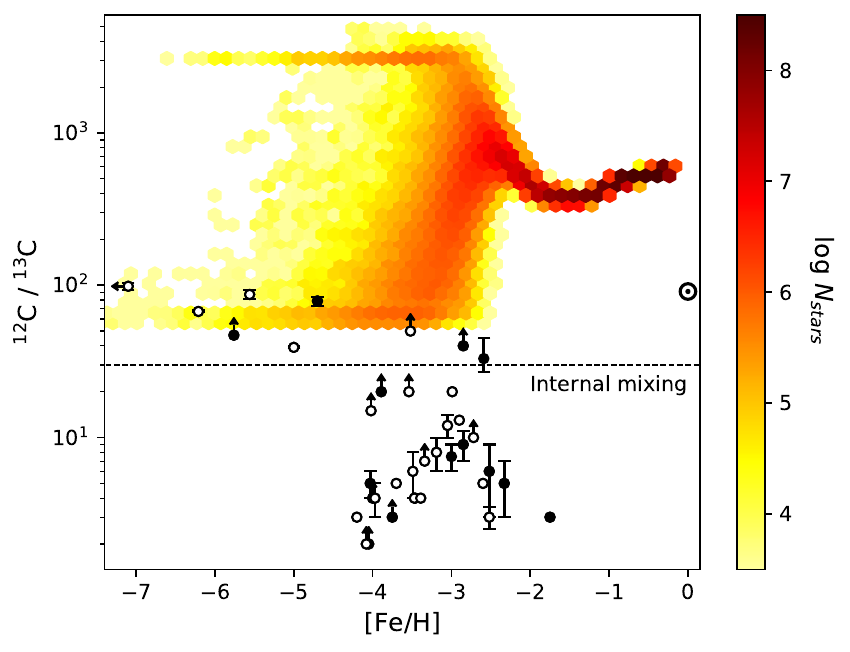}
      \caption{Same model as in Fig.~\ref{fig:SFR} (shell mergers with yields of \citealp{Roberti_2024}), but assuming an IMF from \cite{2001MNRAS.322..231K} instead of \cite{1986FCPh...11....1S}.}
         \label{fig:Kroupa}
   \end{figure}

\begin{figure}
   \centering
   \includegraphics[width=\hsize]{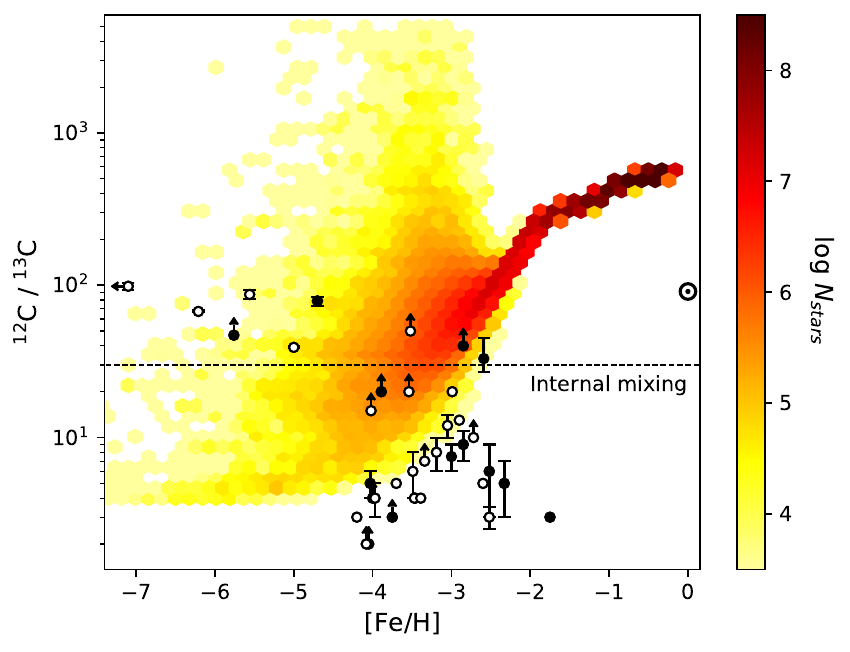}
      \caption{Same model as in Fig.~\ref{fig:SFR0Fe} (shell mergers and outer layer ejection in 20 - 25 M$_\odot$ stars at \feh{\leq-3}), but assuming an IMF from \cite{2001MNRAS.322..231K} instead of \cite{1986FCPh...11....1S}.}
         \label{fig:Kroupa2}
   \end{figure}

The choice of a different initial mass function can potentially have an impact on the number and frequency of H-He shell mergers. We show in Fig.~\ref{fig:Kroupa} and \ref{fig:Kroupa2} the chemical evolution models from Fig.~\ref{fig:SFR} and \ref{fig:SFR0Fe}, respectively, recomputed with the IMF from \cite{2001MNRAS.322..231K} instead of the one by \cite{1986FCPh...11....1S}. Comparing the models with different IMF, it is possible to see that its impact is not significant: the models are slightly shifted towards higher [Fe/H], and at high metallicity \crat{} is larger due to more $^{12}$C and no $^{13}$C coming from massive stars at intermediate metallicity. There is indeed a larger density of stars with lower \crat{}, but the model cannot extend further since the nucleosynthesis sources are the same as before, and the IMF has not a strong enough effect on their number. 


\end{appendix}

\end{document}